\documentclass[referee]{aa}
\usepackage{longtable} 
\usepackage{supertabular}
\usepackage{graphicx}
\usepackage{amssymb}
\usepackage{amsmath}
\usepackage{units}
\usepackage{version}
\usepackage{color}
\usepackage{float}

\reversemarginpar

\def\be{\begin{equation}}
\def\ee{\end{equation}}
\def\bi{\begin{itemize}}
\def\ei{\end{itemize}}

\begin{document}

\title{Joint 3D modelling of the polarized Galactic synchrotron and thermal dust foreground diffuse emission} 
 \author{ L.~Fauvet~\inst{1}\and J.~F.~Mac\'{\i}as-P\'erez ~\inst{1} \and J.~Aumont~\inst{2,3} \and F.X. D\'esert~\inst{1,4,5}  \and T.R.~Jaffe~\inst{3,6} \and A.J.~Banday~\inst{3} \and M.~Tristram~\inst{7}  \and A.H.~Waelkens~\inst{8}  \and D.~Santos~\inst{1}}

\institute{LPSC, Universit\'e Joseph Fourier Grenoble 1, CNRS/IN2P3,
  Institut National Polytechnique de Grenoble, 53 avenue des Martyrs,
  38026 Grenoble cedex, France  \and Institut d'AstrophysiqueSpatiale, Centre Universitaire d'Orsay, 
Bat. 121, 91405  Orsay Cedex, France\and Centre d'Etude Spatiale des
  Rayonnements, 9 avenue du Colonel Roche, 31028 Toulouse, France  
  \and Laboratoire d'astrophysique de
  Grenoble, OSUG, Universit\'e Joseph Fourier BP 53, 38041 Grenoble
  CEDEX 9, France \and Institut Neel, 25 rue des Martyrs, BP 166, 38042
  Grenoble cedex 9, France   \and Jodrell Bank Centre for Astrophysics, School of Physics and Astronomy, The University of Manchester, Oxford Road, Manchester M13 9PL, UK \and  Laboratoire de
l'Acc\'el\'erateur Lin\'eaire, BP 34, 91898 Orsay Cedex
France \and Max-Planck Institute for Astrophysics, Karl Schwarzschild
  Str. 1, 85741 Garching, Germany
 }


\abstract{}{We present for the first time a coherent model of the polarized Galactic synchrotron and thermal
dust emissions which are the main diffuse foreground for the measurement
of the polarized power spectra of the CMB fluctuations with the Planck satellite mission.
}{We produce 3D models of the Galactic magnetic field including regular and
turbulent components, and of the distribution of matter in the Galaxy, relativistic electrons and dust grains. 
By integrating along the line of sight we construct maps of the polarized Galactic synchrotron and
thermal dust emission for each of these models and compare them to currently available data.
We consider the 408 MHz all-sky continuum survey, the  23 GHz band of the Wilkinson Microwave
Anisotropy Probe and the 353 GHz Archeops data.}{ The best-fit parameters obtained are consistent with previous 
estimates in the literature based only on synchrotron emission and pulsar rotation measurements.
They allows us to reproduce
the large scale structures observed on the data. Poorly understood local Galactic structures and turbulence
make difficult an accurate reconstruction of the observations in the Galactic plane.}
{Finally, using the best-fit model we are able to estimate the expected polarized foreground contamination at the Planck frequency
bands. For the CMB bands, 70, 100, 143 and 217~GHz, at high Galactic latitudes although 
the CMB signal dominates in general, a significant foreground contribution
is expected at large angular scales. In particular, this contribution will dominate the CMB signal
for the B modes expected from realistic models of a background of primordial gravitational waves.}


\date{\today}

\titlerunning{Galactic polarized foreground for PLANCK}
\maketitle






\section{Introduction}
\label{introduction}

\indent The PLANCK satellite mission, currently in flight, should permit the
more accurate measurements of the CMB anisotropies both in
temperature and polarization. Planck, which observes the sky on a wide range of
frequency bands from 30 to 857~GHz, has a combined sensitivity of $\frac{\Delta T}{T_{CMB}} \sim 2 \frac{\mu K}{K}$ and
an angular resolution from 33 to 5 arcmin~(\cite{bluebook}).  
Of particular interest is the measurement of the polarization B modes which implies the presence of
tensor fluctuations from primordial gravitational waves generated during inflation.
Planck should be able to measure the tensor-to-scalar ratio, $r$, down to 0.1  (\cite{betoules2009,efstathiou1}) in the case
of a nominal mission (2 full-sky surveys) and to 0.05 for the Extended Planck Mission: four
full-sky surveys (\cite{efstathiou2}) . The value of $r$
sets the energy scale of the inflation (\cite{peiris}) and then provides
constraints on inflationnary models (\cite{baumann}).\\ \\
\indent To achieve this high level of sensitiviy it is necessary to
accurately estimate the temperature and polarization foregrounds mainly from diffuse Galactic
emission components -- synchrotron, thermal and rotational dust, and free-free --  as well as from the Galactic and extra-Galactic
point-like and compact sources. Indeed, at the Planck frequency bands these foreground components may dominate the CMB 
signal and therefore, they need to be either masked or subtracted prior to any CMB analysis. For this purpose,
the Planck collaboration plans to use component separation techniques (see \cite{leach} for a summary) in addition 
to the traditional masking of highly contaminated sky regions including identified point-like and
compact sources. As these component separation techniques will be mainly based
on Planck data only, one of the main issues will be to estimate the residual foreground
contamination on the final CMB temperature and polarization maps. These residuals
will translate into systematic biases and larger error bars on the estimation of the temperature
and polarization power spectra of the CMB fluctuations (see \cite{betoules2009} for a recent study). 
Thus, they will impact the precission to which 
cosmological information can be retrieved from the Planck data. \\ \\

\indent For polarization the main foreground contributions will come from the diffuse Galactic synchrotron 
and thermal dust emission. From the WMAP observations, \cite{page2007} have shown 
that the radio synchrotron emission from relativistic electrons is highly polarized, up to 70~\%, between 23 and 94GHz.
Furthermore, \cite{benoit2004a,ponthieu2005} have observed signicantly polarized thermal dust emission, up 
to 15 \% at the 353~GHz Archeops channel. By contrast the diffuse free-free emision 
is not polarized and the anomalous microwave emission has been measured to be weakly polarized, $3 ^{+1.3}_{-1.9}$ \%,  
(\cite{battistelli2006}). Finally, at the Planck frequency bands the polarized contribution from compact and point
sources is expected to be weak for radio (\cite{nolta2009}) and dust (\cite{desert2008}) sources.
The spatial and frequency distribution of both Galactic synchrotron and thermal dust polarized emission
at the Planck frequencies are not well known and the only available information comes from micowave
and submillimetre observations. For synchrotron, Faraday rotation (\cite{burn1966}) makes it very difficult to extrapolate
the polarized observed radio emission (\cite{wolleben,wolleben07, carretti2009}) to the microwave domain. 
For thermal dust, polarized observations are not currently available in the infrared and the current optical 
data (\cite{heiles}) are too sparse (\cite{page2007}) for a reliable extrapolation to lower frequencies.
\\

\indent{The diffuse Galactic synchrotron emission is produced by relativistic electrons spiraling around
the Galactic magnetic field lines and its polarization is orthogonal both to the line-of-sight and
to the field direction (\cite{ribicki}). Based on these statements, \cite{page2007} proposed to model the polarized synchrotron 
Galactic emission observed by the WMAP satellite using a 3D model of the Galaxy
including the distribution of relativistic electrons and the Galactic magnetic field structure. 
Although this model allowed them to explain the observed polarization angle at the 23~GHz band
where the synchrotron emission dominates, it was not used for the CMB analysis. Instead they constructed a template
of the polarized synchrotron emission from the 23 GHz band and extrapolated it to higher frequencies.
Independently, \cite{han2004,han2006} has also proposed a 3D model of the Galactic free electrons (\cite{cordes}) and
of the Galactic magnetic field including a regular and a turbulent component to explain the observed rotation measurements
on known pulsars. Based on previous works \cite{sun} has performed a combined analysis of the polarized WMAP data and
of the rotation measurements of pulsars using the publicly available HAMMURABI code (\cite{waelkens})
for computing the integrated emission along th line-of-sight.
This work has been extended by \cite{jaffe} for the study of the Galactic plane using a MCMC algorithm
for the determination of the parameter of the models and by \cite{jansson} for the full sky
using a likelihood analysis for parameter estimation. \\



\indent Dust grains in the Interstellar Medium (ISM) are heated by stellar radiation and radiate
in the form of thermal dust emission (\cite{desert1998}).
They are considered to be oblate and to align with their longitudinal axis perpendicular to
the magnetic field lines (\cite{davis}). When aligned they will end up rotating with their angular
moment parallel to the magnetic field direction.
As the thermal dust emission is more efficient along the long axis, 
this generates linear polarization orthogonal to the magnetic field direction and to the line-of-sight. 
The polarization fraction of the emission depends on the distribution of the size of the grains and is about a few percent 
at the millimeter wavelengths (\cite{hildebrand1999, vaillancourt}).
\cite{ponthieu2005} concluded that the polarized emission observed in the 353~GHz Archeops data was associated
with the thermal dust emission and proposed a simple magnetic field pattern to explain the measured direction
of polarization on the Galactic plane. \cite{page2007} suggested that part of the observed polarized emission 
at the 94~GHz WMAP data was also due to thermal dust. They modeled it using the observed polarization direction
of stellar light (\cite{heiles}) which from the above statement must be perpendicular to that of thermal dust. \\

\indent With the prospect of the data analysis of the Planck satellite mission in mind,
we present here consistent physical models of the synchrotron and thermal
dust emission based on the 3D distribution of relativistic electrons and dust
grains on the Galaxy, and on the 3D pattern of the Galactic magnetic field.
The paper is structured as follows: Sect.~\ref{data} describes the 408 MHz all-sky continnum survey
(\cite{haslam}), the five-year WMAP data set (\cite{page2007}) and
the Archeops data (\cite{ponthieu2005}) used in the analysis.  In
Sect.\ref{3dgal_model} we describe in detail the models. Section~\ref{gal_comp}
describes the statistical comparison of data and models.  In Sect.~\ref{gal_bias}
we discuss the impact of polarized foreground emission on the measurement of
the polarized CMB emission with the Planck satellite. Finally, conclusions are presented in Section
\ref{conc}.

\section{Observational data}
\label{data}

\subsection{Diffuse Galactic synchrotron emission}

The Galactic diffuse synchrotron emission is important at radio and microwave observation frequencies.
Although its SED is not accurately known it can be well represented by a power law in antena temperature
$I_{\nu} \propto \nu^\beta_s$ with the synchrotron spectral index ranging from $-2.7$ to $-3.3$ (\cite{kogut,gold}).
Radio frequency surveys as the Leider 408 MHz and 1.4 GHz surveys (\cite{brouw,wolleben}), the Parkes survey at 2.4 GHz (\cite{duncan1999}), 
and the MGLS survey (\emph{Medium Galactic Latitude Survey}) at 1.4 GHz (\cite{uyaniker}) are generally used to trace the Galactic
diffuse synchrotron emission in intensity. For polarization Faraday rotation makes things more complex as for frequencies lower than
10~GHz strong depolarization is expected (\cite{burn1966,sun,jaffe, jansson}) so that the best Galactic diffuse synchrotron tracers
are at high frequency like the WMAP survey at 23~GHz (\cite{page2007}).

\subsubsection{408 MHz all-sky continuum survey}
\label{data_has}
\begin{figure*}
\centering
\includegraphics[angle=90,height=3.5cm,width=7cm]{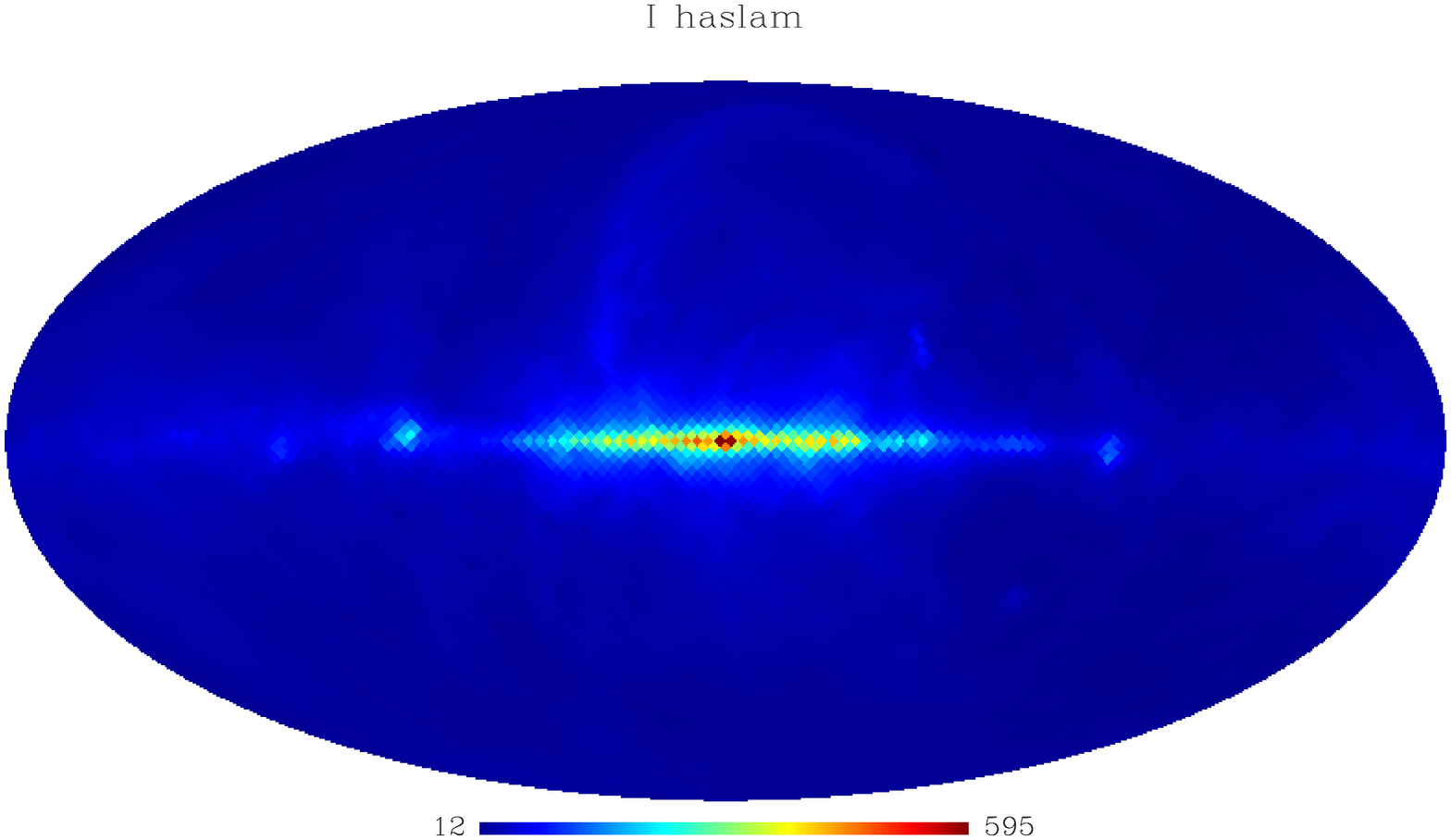}\includegraphics[angle=90,height=3.5cm,width=7cm]{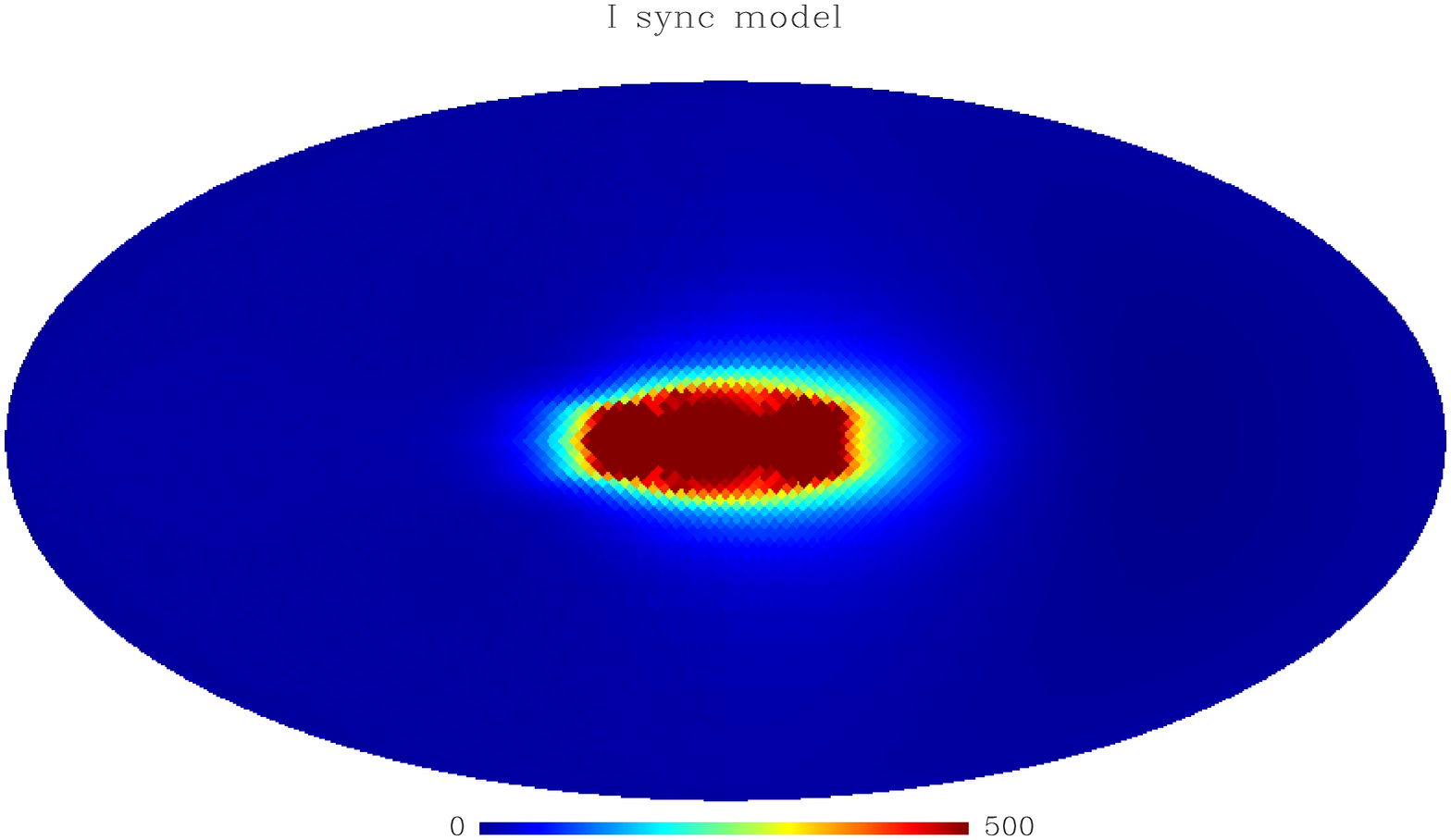}
\caption{Intensity maps at 408 MHz in K$_{RJ}$ units for the Haslam data \emph{(left)} and built with the mode of synchrotron emission with MLS  magnetic field for the best fit model parameters \emph{(right)}. \label{map_has_Isync_gp}}
\end{figure*}

\indent In the following we use the 408 MHz all-sky
continuum survey (\cite{haslam}), after subtraction of the free-free emission at this frequency,
as a tracer of the Galactic synchrotron emisison in temperature. We use the  HEALpix (\cite{gorski}) format map
available on the LAMBDA website \footnote{http://lambda.gsfc.nasa.gov/}. The calibration scale of this survey is claimed to be accurate to better
than 10 \% and the average zero level has an uncertainty of $\pm 3$ K as
explained in \cite{haslam}. To substract the free-free emission at 408 MHz we 
use the five-year public WMAP (\emph{Wilkinson Microwave Anisotropies Probe}) free-free foreground map at 23~GHz generated from the maximum
entropy method (MEM) described in \cite{hinshaw}. We have found that the free-free correction has no impact on the final results presented on this paper. We start from the full-sky HEALpix maps at
$N_{side}=512$ (pixel size of 6.9 arcmin) and downgrade them to $N_{side} = 32$ (pixel size of
27.5 arcmin). We then substract from the Haslam data the free-free
component extrapolated from the K-band assuming a power-law dependence
of $\nu^{-2.1}$ as in \cite{dickinson}. The left pannel of Figure~\ref{map_has_Isync_gp} shows the free-free corrected 408 MHz all-sky survey where we clearly observe the Galactic plane and the North Celestial Spur at high Galactic latitude.

\subsubsection{Five-year WMAP polarized data at 23 GHz}

\begin{figure*}
\centering
\includegraphics[angle=90,height=3.5cm,width=7cm]{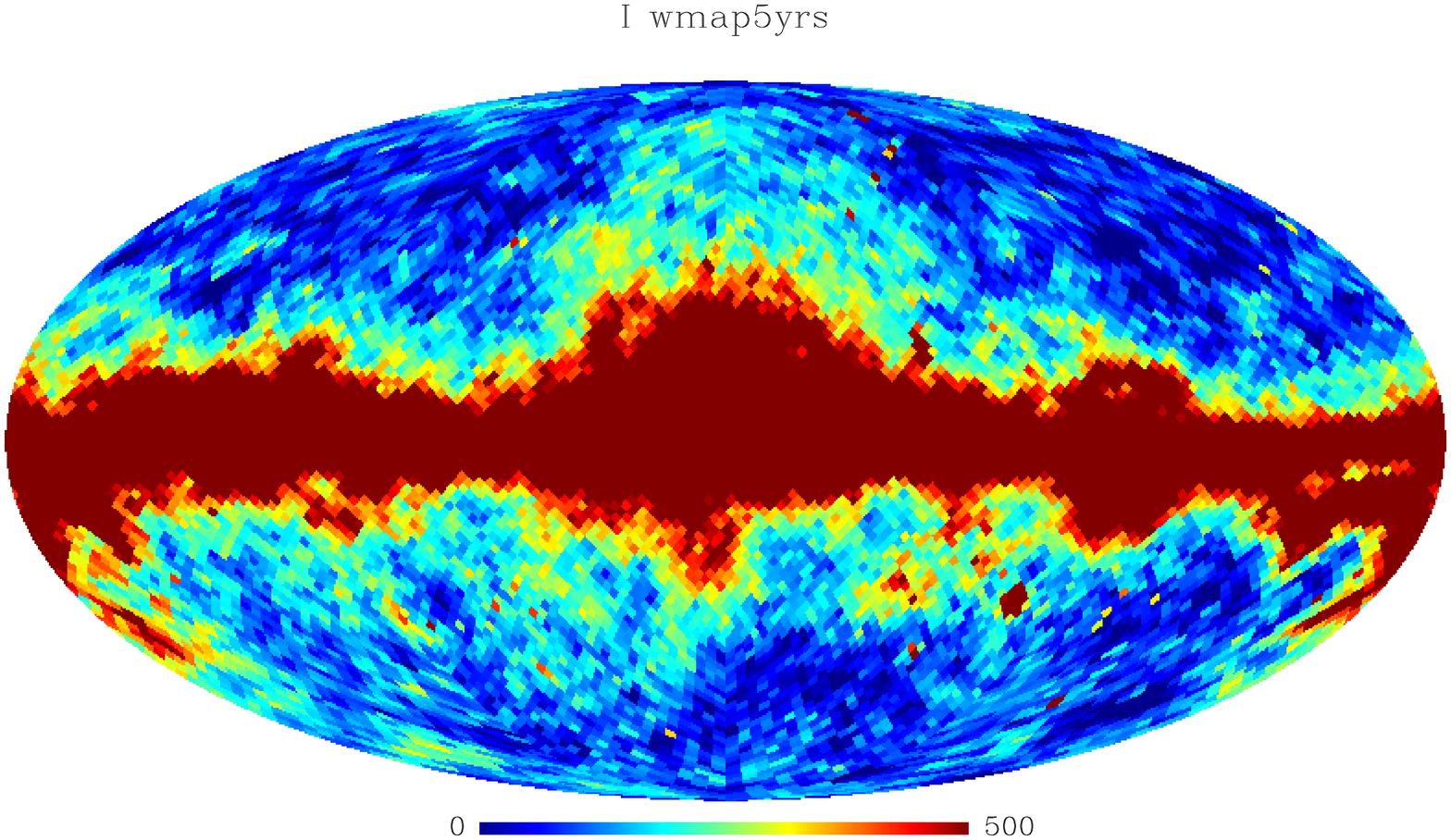}\includegraphics[angle=90,height=3.5cm,width=7cm]{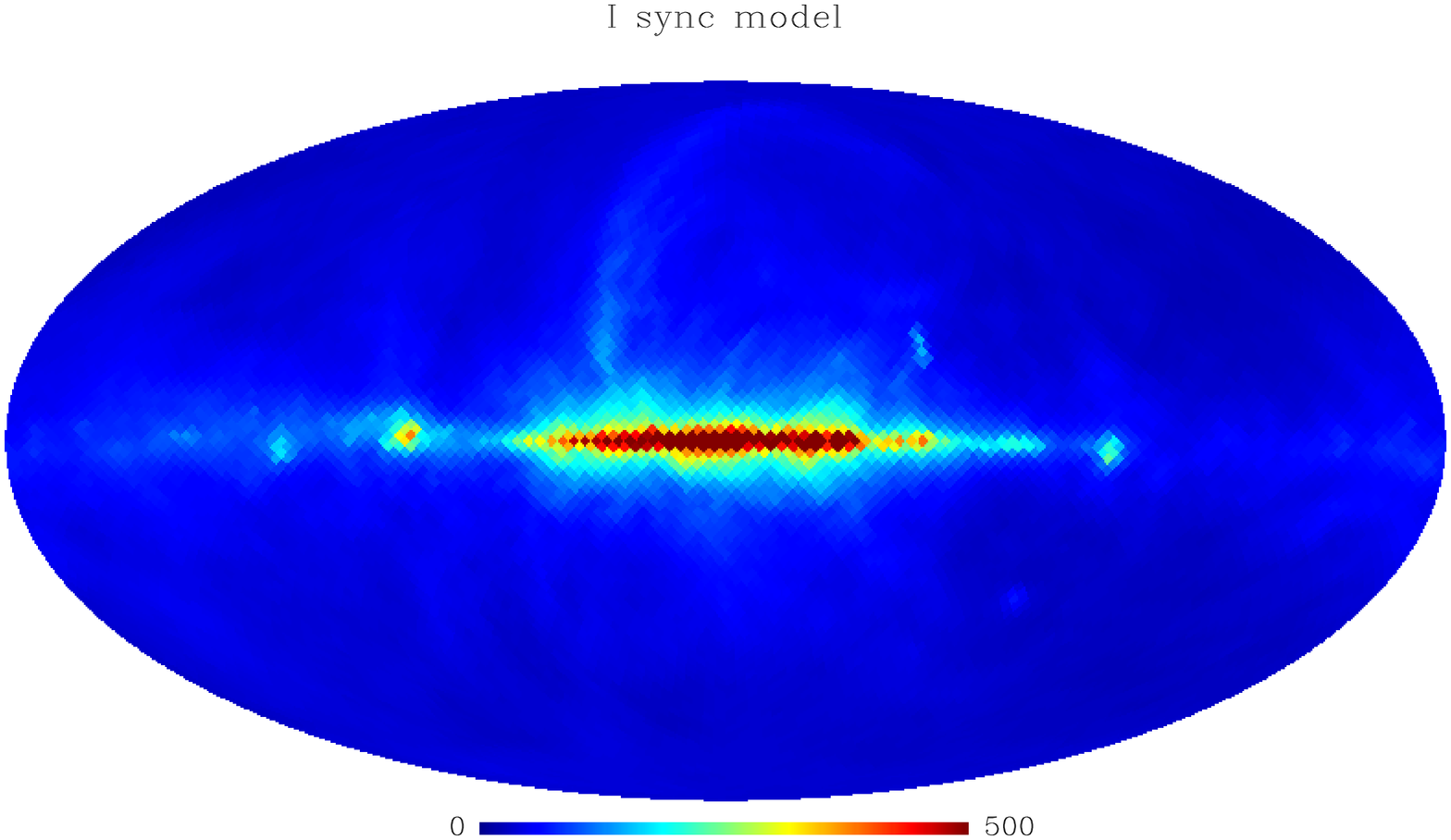}
\includegraphics[angle=90,height=3.5cm,width=7cm]{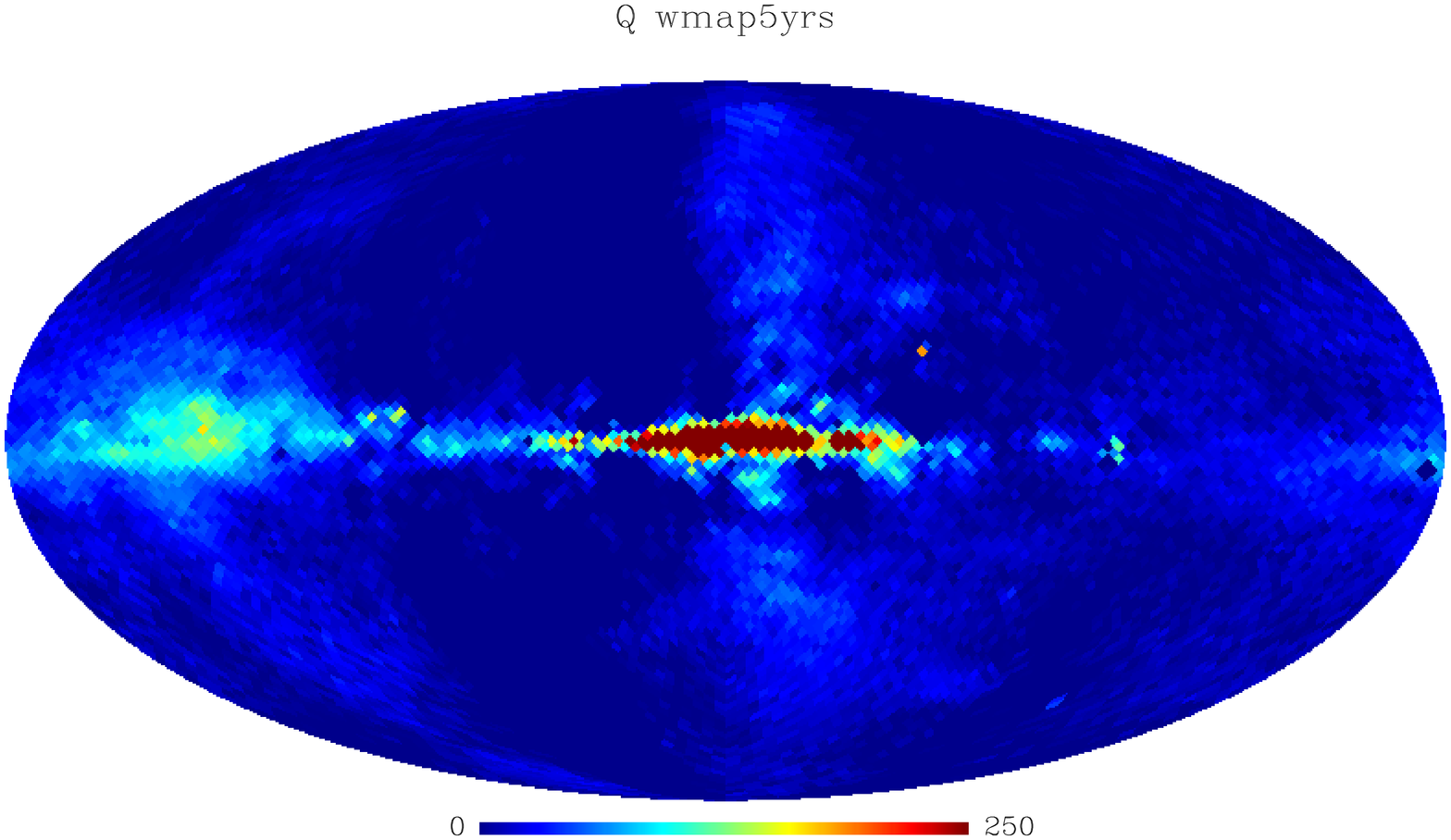}\includegraphics[angle=90,height=3.5cm,width=7cm]{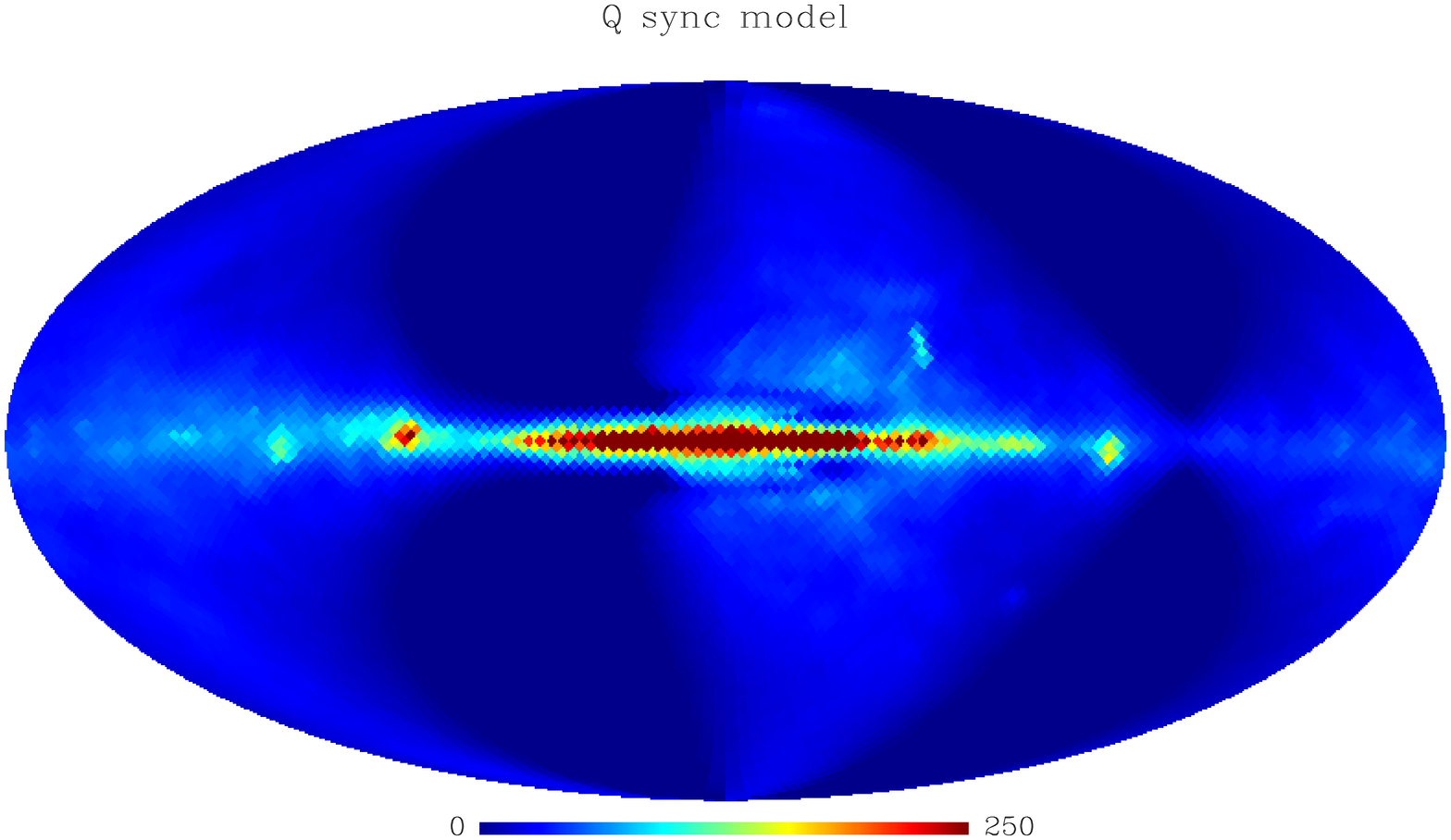}
\includegraphics[angle=90,height=3.5cm,width=7cm]{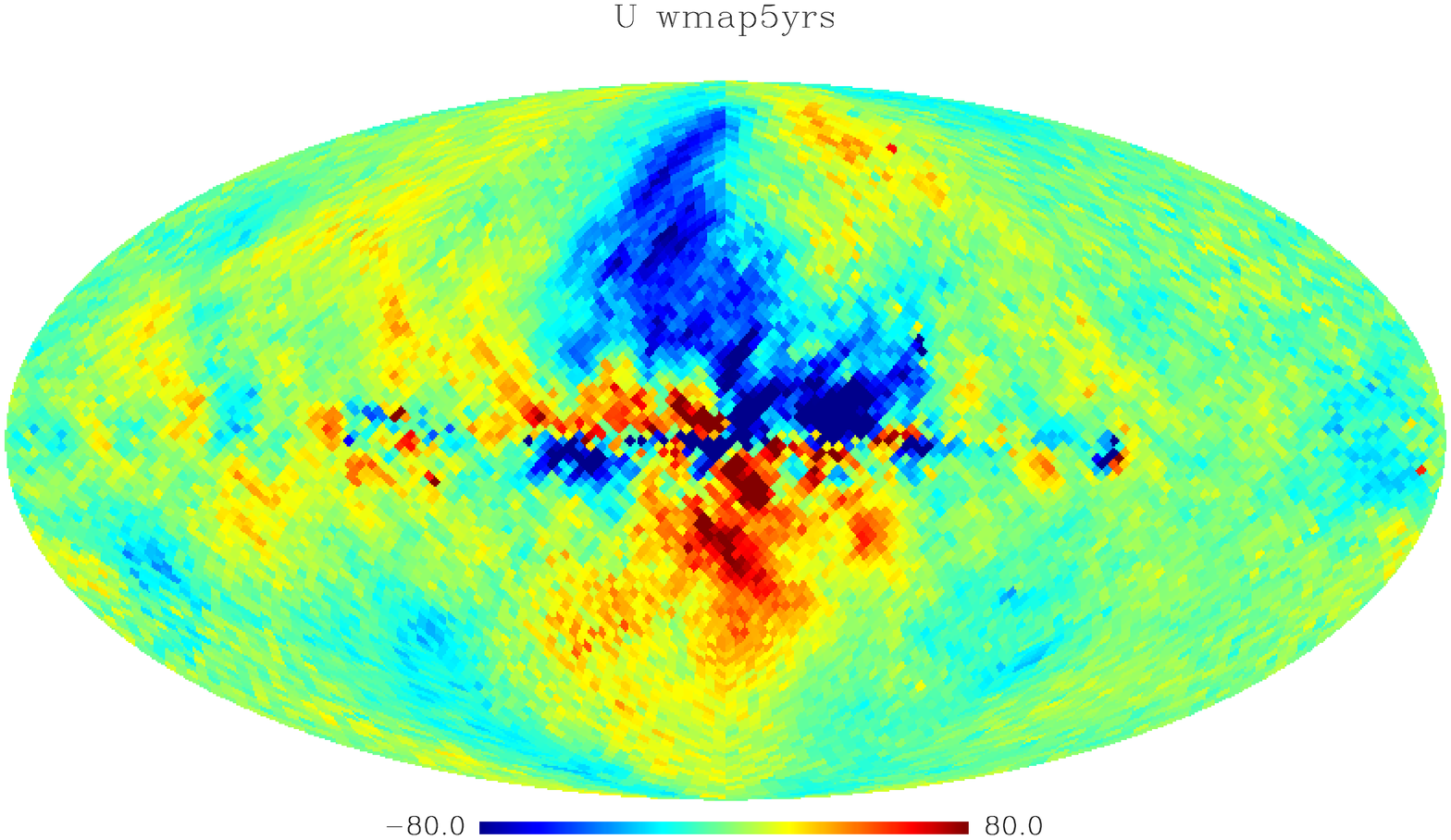}\includegraphics[angle=90,height=3.5cm,width=7cm]{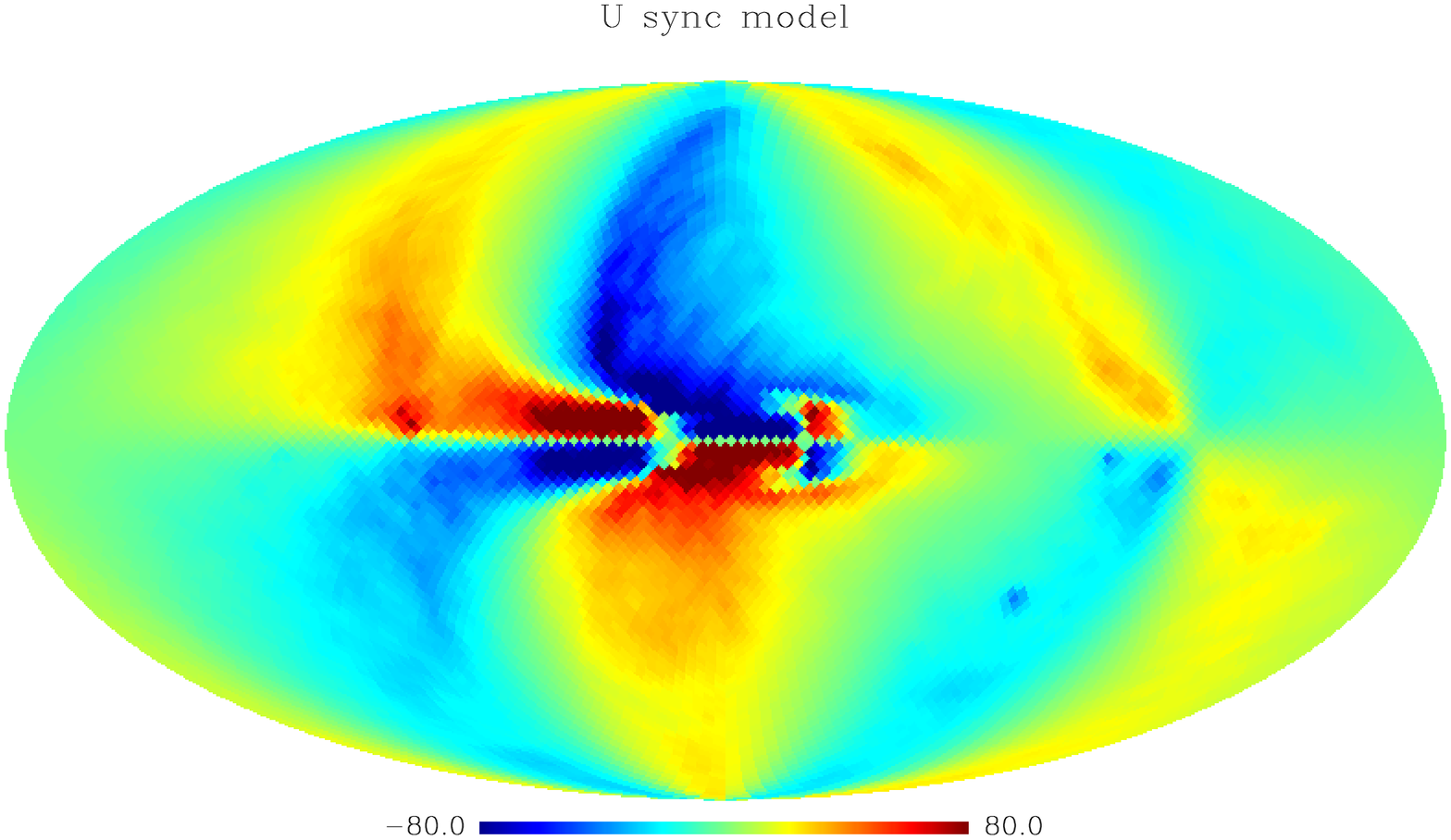}
\includegraphics[angle=90,height=3.5cm,width=7cm]{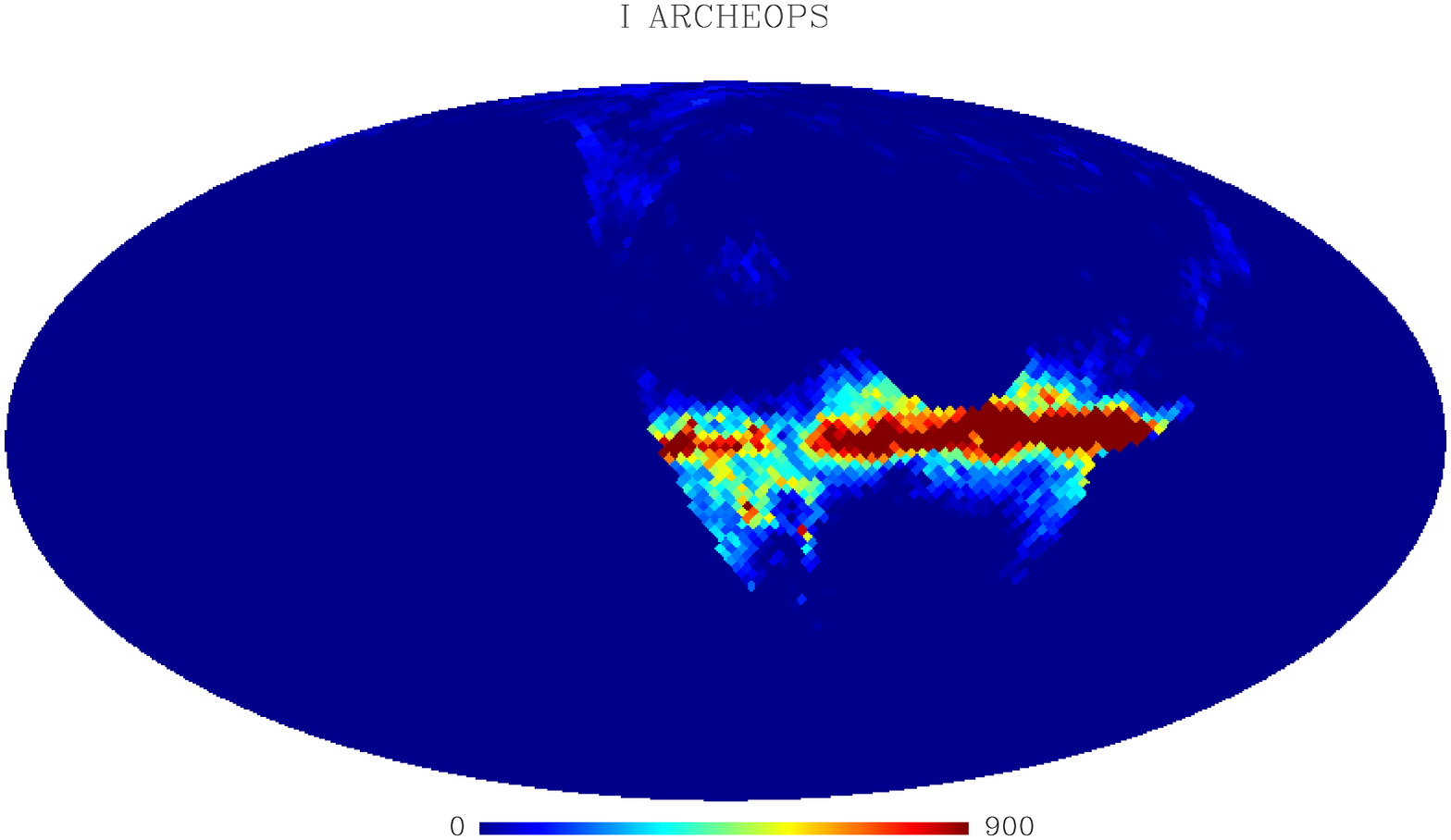}\includegraphics[angle=90,height=3.5cm,width=7cm]{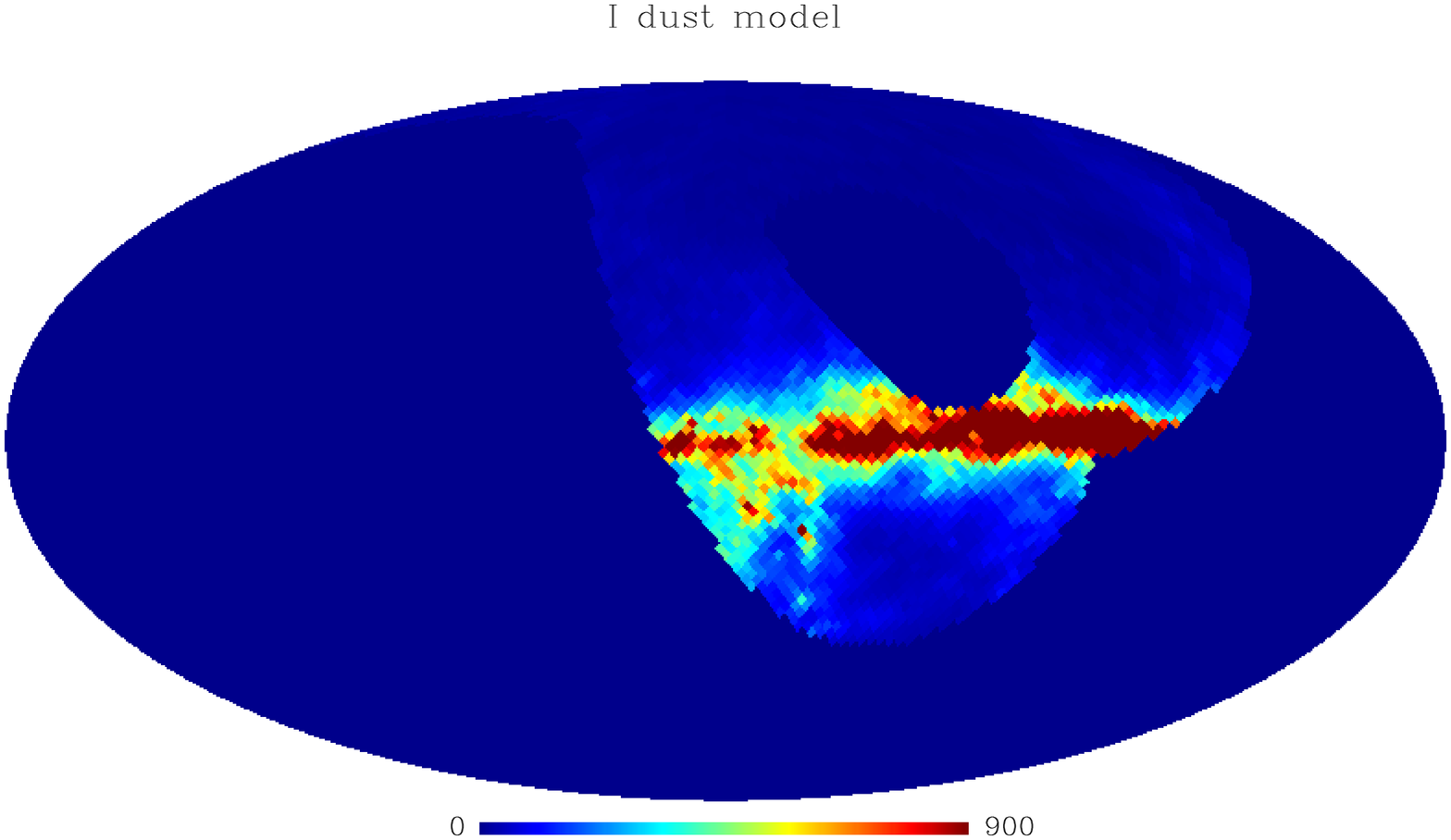}
\includegraphics[angle=90,height=3.5cm,width=7cm]{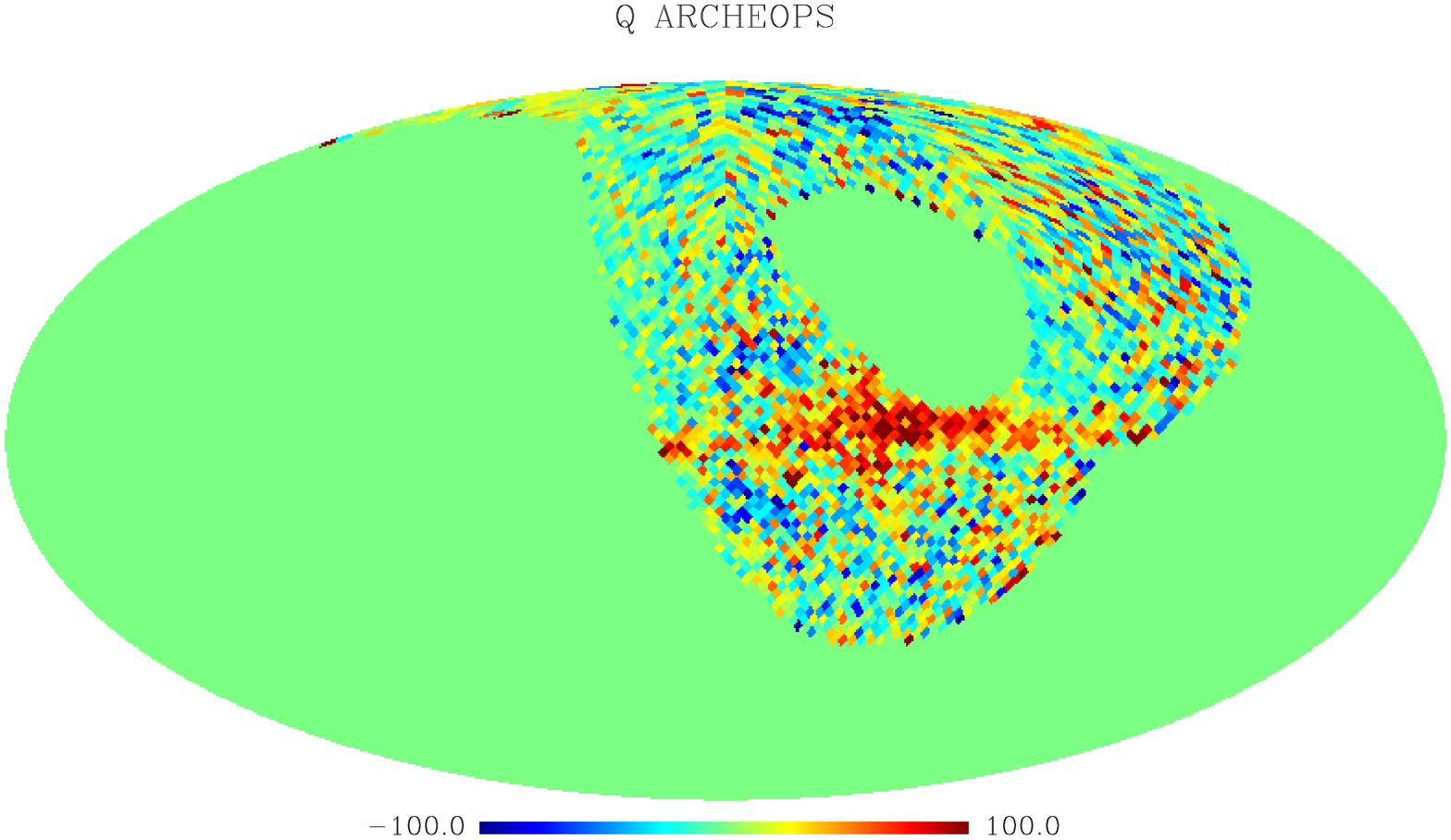}\includegraphics[angle=90,height=3.5cm,width=7cm]{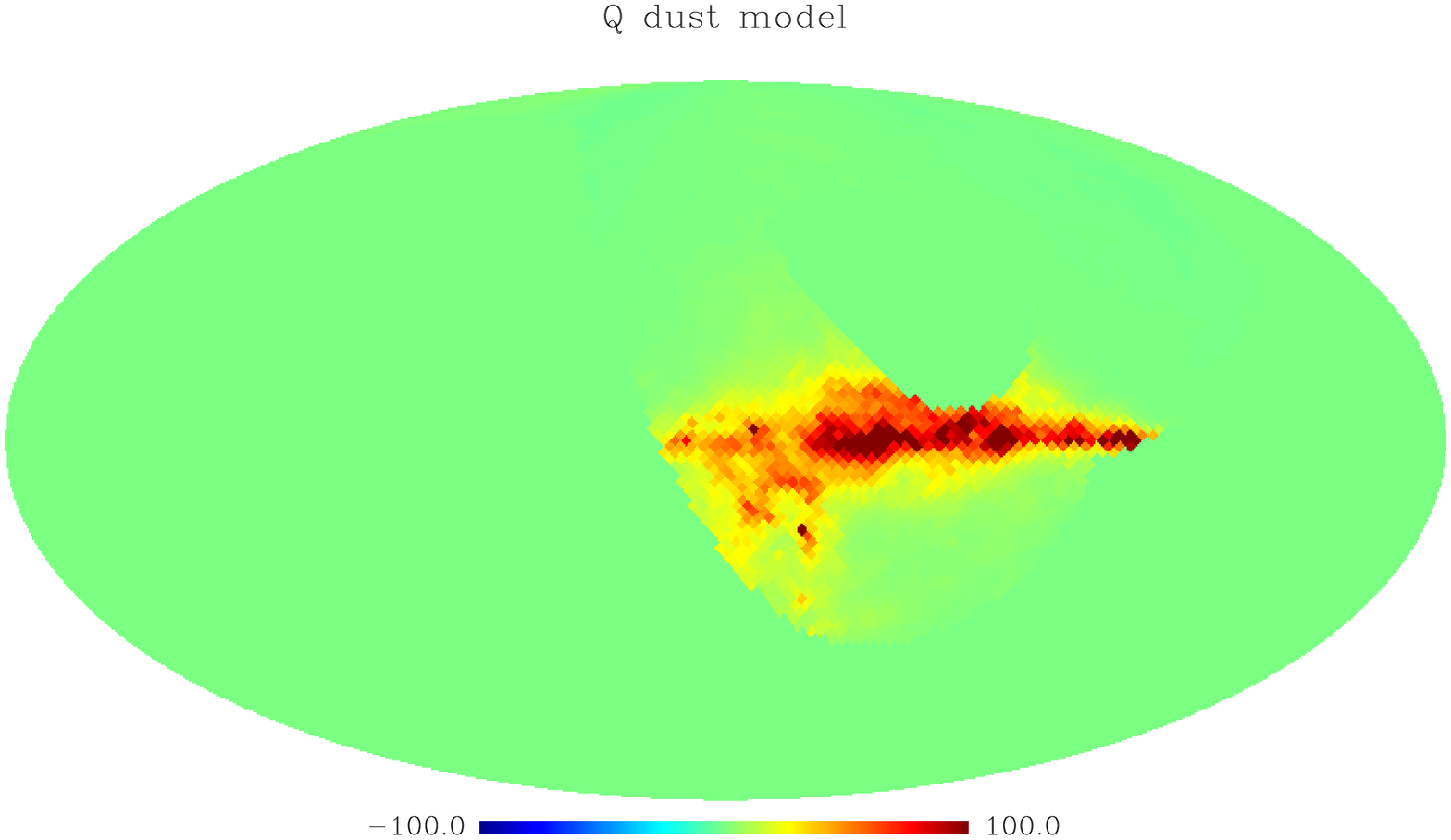}
\includegraphics[angle=90,height=3.5cm,width=7cm]{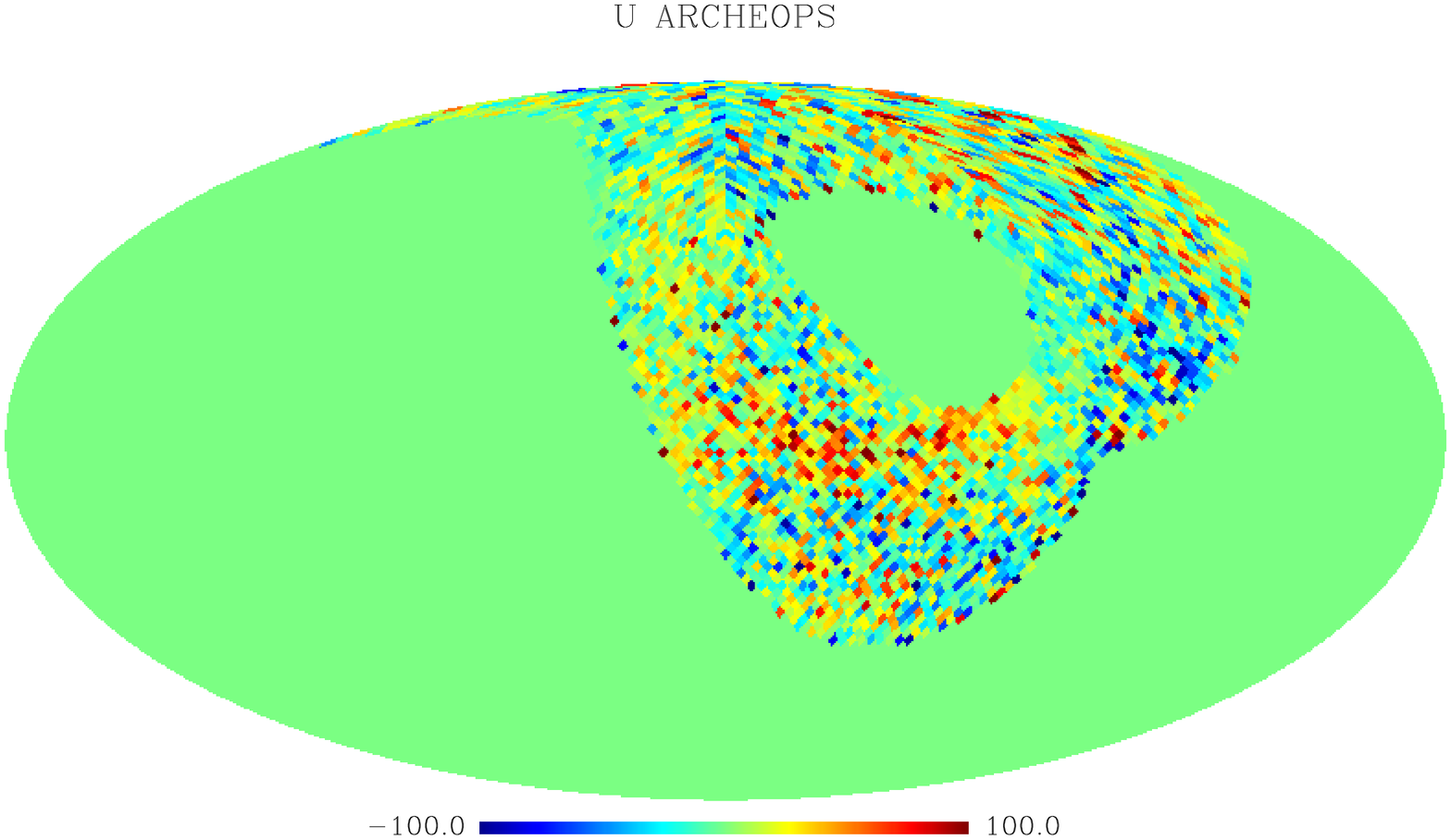}\includegraphics[angle=90,height=3.5cm,width=7cm]{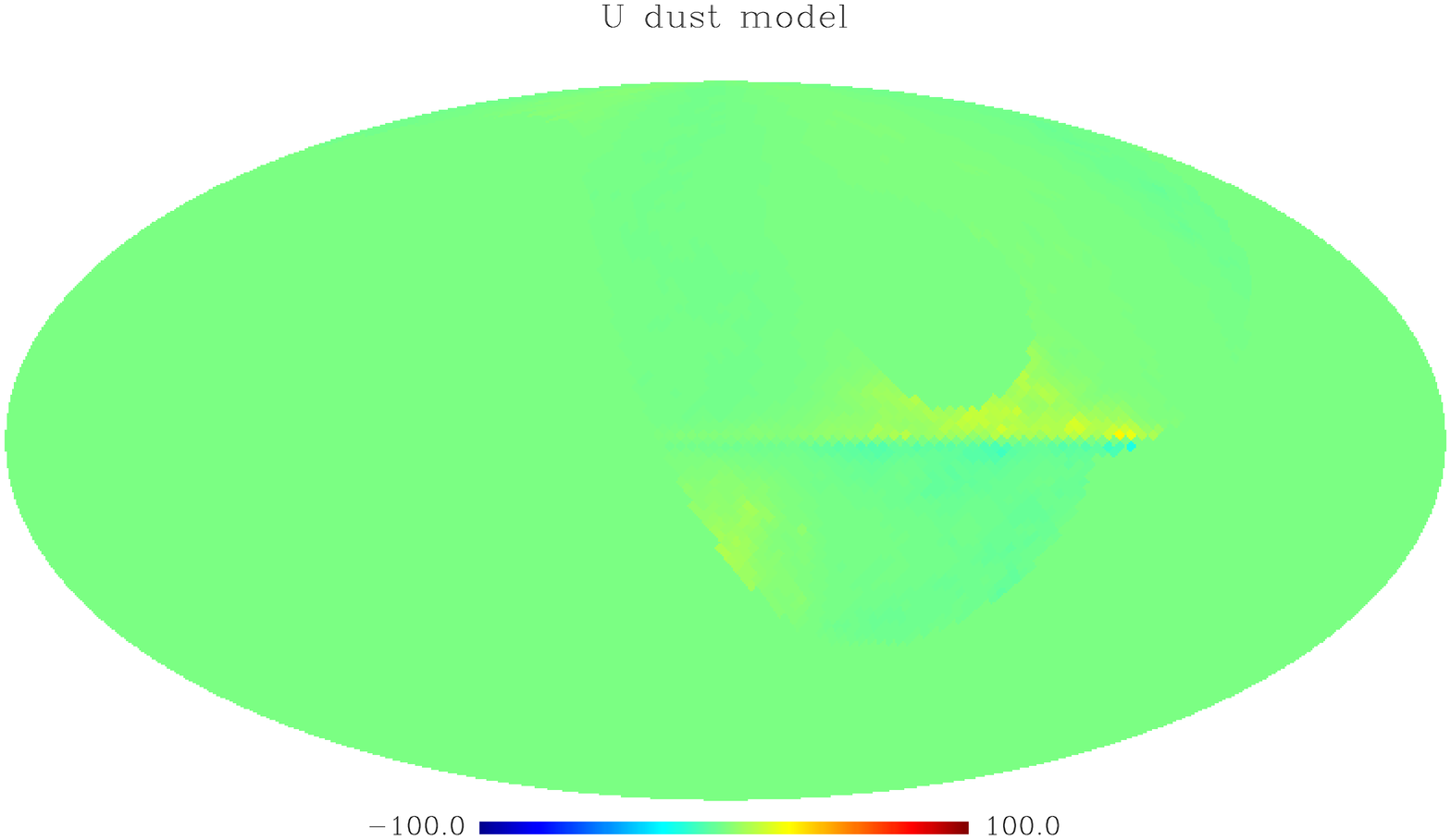}\caption{\emph{Form top to bottom:} Maps in intensity, I, and polarization Q and U at 23 GHz for the WMAP 5-year data \emph{(left)} and the model of synchrotron emission with MLS magnetic field for the best fit model parameters \emph{(right)} and at 353 GHz for the Archeops data \emph{(left)} and the model of thermal dust emission with MLS magnetic field for the best fit model parameters \emph{(right)}. The 353~GHz maps are rotated by 180$^{\circ}$ for better visualization. All the maps are in $K_{RJ}$ units.\label{map_wmap_arch}}
\end{figure*}

\indent To trace the polarized synchrotron emission we use the all-sky five-year WMAP Q and U maps at 23~GHz available
on the LAMBDA website in the
HEALPix pixelisation scheme (\cite{gorski}). In particular the coadded maps at $N_{side}=512$ for the K-band
(\cite{page2003,gold}). These maps are then downgraded to $N_{side} = 32$ to increase the signal-to-noise ratio
as we are only interested on very large angular scales and the analysis will be performed on Galactic
latitude profiles. The pixel window function will be accounted for in the following. 
We assume anisotropic white noise on the maps and compute the variance per pixel using the variance per hit
provided on the LAMBDA website and the maps of number of hits. The large angular scale correlations on the
noise are neglected. We think this has no impact on the final results as very similar results have been
obtained using a pixel-to-pixel analysis at $N_{side} = 16$ using the noise correlation matrix. 
The second and third plots on the left column of Figure~\ref{map_wmap_arch}
show the 23~GH Q and U maps. We can clearly observe the Galactic plane but also large-scale high Galactic structures.

\subsection{Thermal dust}
\indent The dust thermal emission in temperature is well traced by the IRAS (\cite{schlegel}) all-sky observations in the infrared, the FIRAS-COBE (\cite{boulanger})  all-sky observations in the radio and millimeter domain and the ARCHEOPS (\cite{macias, benoit2004a}) data in the millimeter domain over 30 \% of the sky .
\indent In polarization early observations by \cite{hiltner,hall} and later by \cite{heiles} have proved that the starlight emission in the
optical domain was polarized and therefore we will expect the thermal dust emission at millimetric wavelength also to be polarized. 
This was confirmed by the Archeops observations at 353~GHz (\cite{ponthieu2005}) that yielded a polarization fraction of about 10 \% in the Galactic plane.
 
\indent Here  we use the Archeops 353 GHz  Q and U maps as tracers of the polarized thermal dust emission.
As shown in the fifth and sixth plots of  the left column of Figure~\ref{map_wmap_arch} they cover about 30 \% of the sky with 13 arcmin resolution.
In contrast to the WMAP data at 23~GHz, most the signal in the Q and U 353~GHz maps is concentrated on the Galactic plane.
These maps are then downgraded to $N_{side} = 32$ to increase the signal-to-noise ratio. As for the WMAP 23~GHz data
the noise assumed to be anisotropic white noise on the maps and compute the variance per pixel using the variance
provided by the Archeops collaboration (\cite{macias}).

\section{3D modeling of the galaxy}
\label{3dgal_model}

\begin{figure*}
\centering
\includegraphics[scale=0.35]{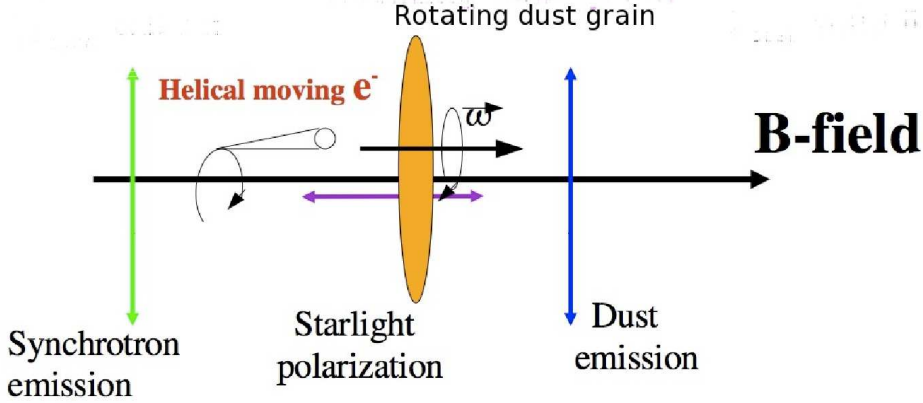}\caption{Schematic view of the 
direction of polarization of the Galactic synchrotron and dust thermal emission as a function of the Galactic magnetic
field direction. \label{syncdustpol}}
\end{figure*}

\indent We present in this section a realistic model of the diffuse synchrotron and dust polarized emission
using a 3D model of the Galactic magnetic field and of the matter density in the Galaxy. We will
consider the distribution of relativistic cosmic-ray electrons (CREs), $n_\mathrm{CRE}$, for the synchrotron emission and the distribution of dust grains, $n_{dust}$, for the dust thermal emission. The total polarized foreground emission observed at a given position on the 
sky $\mathrm{{\bf n}}$ and at a frequency $\nu$ can be computed by integrating along the line of sight as follows.\\

\subsubsection*{Synchrotron}
For the synchrotron emission (\cite{ribicki}) and defining 
\begin{eqnarray}
\mathrm{d}I^{sync}_{\nu} = \epsilon^{sync} (\nu) \ \ n_\mathrm{CRE}(\mathrm{{\bf n}},z)  \\  \nonumber
. \left(B_l(\mathrm{{\bf n}},z)^2 + B_t(\mathrm{{\bf n}},z)^2 \right )^{(s+1)/4} dz
\end{eqnarray}
we obtain
\begin{eqnarray}
\centering
I^{sync}_{\nu} (\mathrm{{\bf n}})  &=& \int \mathrm{d}I^{sync}_{\nu}, \\
Q^{sync}_{\nu} (\mathrm{{\bf n}}) &=& \int \! \! \! \mathrm{d}I^{sync}_{\nu} \cos(2\gamma(\mathrm{{\bf n}},z)) \ p^{sync}, \\
U^{sync}_{\nu} (\mathrm{{\bf n}}) &=& \int \mathrm{d}I^{sync}_{\nu} \sin(2\gamma(\mathrm{{\bf n}},z)) \ p^{sync}.
\end{eqnarray}
\label{eq_ms}

\noindent where I, Q and U are the Stokes parameters and $\epsilon^{sync} (\nu)$ is an emissivity term.
$\gamma$ is the polarization angle and
$B_n$, $B_l$ and $B_t$ are the magnetic field components along,
longitudinal and transverse to the line of sight, ${\bf n}$, and $z$ is a 1D coordinate along the line-of-sight.
$s$ is the exponent of the power-law representing the energy distribution of relativistic
electrons in the Galaxy. The polarization fraction, $p^{sync}$, 
is related to $s$, as follows
\be
\centering
p^{sync} = \frac{s + 1}{s + 7/3}
\ee
\noindent In the following we will assume a constant value of $3$ for $s$ so that
the synchrotron emission will be proportional to the square of the perpendicular component
of the Galactic magnetic field to the line of sight and $p_{sync} = 0.75$ (\cite{ribicki}).\\

Locally, the direction of polarization will be orthogonal to the magnetic field lines and to the line of sight. Then, the polarization angle $\gamma $ is given by 
\begin{eqnarray}
\centering
 \gamma(\mathrm{{\bf n}},s) &=& \frac{1}{2} \arctan{\left( \frac{2B_l(\mathrm{{\bf n}},z) \cdot B_t(\mathrm{{\bf n}},z)}{B^2_l(\mathrm{{\bf n}},z) -B^2_t(\mathrm{{\bf n}},z)} \right)}.
\end{eqnarray}

\subsubsection*{Thermal dust}
\noindent For thermal dust emission and defining
\be
\centering
\mathrm{d}I^{dust}_{\nu} (\mathrm{{\bf n}}) =  \epsilon^{dust} (\nu) \ \ n_{dust}(\mathrm{{\bf n}},z) \ dz
\ee
we can write
\begin{eqnarray}
\centering
I^{dust}_{\nu} (\mathrm{{\bf n}}) &=& \int \mathrm{d}I^{dust}_{\nu}, \\
Q^{dust}_{\nu} (\mathrm{{\bf n}}) &=& \int \mathrm{d}I^{dust}_{\nu}   p^{dust} \cos(2 \gamma(\mathrm{{\bf n},z})) \nonumber \\ && . f_{\mathrm{g}}(\mathrm{{\bf n}},z)  f_{\mathrm{ma}}(\mathrm{{\bf n}},z), \\
U^{dust}_{\nu}(\mathrm{{\bf n}}) &=& \int \mathrm{d}I^{dust}_{\nu}   p^{dust} \sin(2 \gamma(\mathrm{{\bf n}},z))  \nonumber \\ && . f_{\mathrm{g}}(\mathrm{{\bf n}},z)  f_{\mathrm{ma}}(\mathrm{{\bf n}},z).
\end{eqnarray}
where $\epsilon^{dust}$ is the dust emissivity, $p^{dust}$ is the polarization fraction, $\gamma$ is the polarization angle, and, $f_{\mathrm{g}}$ and $f_{\mathrm{ma}}$
are polarization suppression factors (see below). Hereafter, the polarization fraction $p^{dust}$ is set to 10\% following \cite{ponthieu2005}.
\\

As discussed before, dust grains in the ISM are oblate and will align with their large axis (see Figure~\ref{syncdustpol})
perpendicularly to the magnetic field lines (\cite{davis, lazarian1995, lazarian1997, lazarian2009}). Therefore, the polarization direction, for
thermal dust emission will be perpendicular both to the magnetic field lines and the line-of-sight
as was already the case for the synchrotron emission. Then, the polarization angle $\gamma$ will be the same
for the synchrotron and dust emission. However, as the dust grains rotate with their spin axis
parallel to the magnetic field we also need to account for a geometrical supression factor. For example,
if the magnetic field direction is parallel to the line-of-sight we expect the dust polarized emission to be
fully supressed. The suppression factor can be expressed as $f_{\mathrm{g}} = \sin^2(\alpha)$ 
where $\alpha$ is the angle between the magnetic field direction and the line-of-sight. By construction, we observe
that $\gamma$ and $\alpha$ are the same angle. The process of alignement of the dust grains with the
magnetic field is very complex (\cite{mathis1986,goodman1995,lazarian1995, lazarian1997, lazarian2009}) and its accurate representation is out of the scope of this paper.
Then, to account for missalignements between the dust grains and the magnetic field lines we define an empirical factor
$f_{ma}$. The form of this factor is unknown but we have empirically observed that the geometrical supression seems to be
more important than expected for the Archeops data. Therefore, we have taken $f_{ma}$ to be $\propto \sin(\alpha)$.
We have observed that the results presented on this paper are robust with respect to this parameter.

\subsection{Matter density model}

\indent In galactocentric cylindrical coordinates $(r,z,\phi)$ the relativistic electron density distribution
can be written as (see \cite{drimmel})
\begin{eqnarray}
\centering
 n_\mathrm{CRE}(r,z) &=& n_{0,e} \cdot \frac{e^{-\frac{r}{n_{\mathrm{CRE},r}}}}{\cosh(z/\unit[]{kpc})\cosh(z/n_{\mathrm{CRE},h})},
\end{eqnarray}
where $n_{\mathrm{CRE},h}$ defines the width of the distribution vertically and is setted to $1$ kpc in the following. $n_{\mathrm{CRE},r}$ defines the distribution radially and it is a free parameter of the model. Notice that we expect these two paramters to strongly correlated and that is why we have decided to fix one of them as was the case on previous analyses
(\cite{sun,jaffe}).\\

\noindent  The density distribution of dust grains in the Galaxy is poorly known and that is why
we have chosen to describe it in the same way that the relativistic electrons one

\be
\centering n_d(r,z) = n_{0,d} \cdot \frac{e^{-\frac{r}{n_{d,r}}}}{\cosh^2(z/n_{d,h})},
\ee
where $n_{d,r}$ and $n_{d,h}$ are the radial and vertical widths of the distribution.
In the following we set them to 3 and 0.1 kpc respectively.

\subsection{Galactic magnetic field model}
\label{mg_field_model}

\indent According to observations. many spiral galaxies in our vicinity and for various redshifts, present a large scale
magnetic field with intensity of the order of few $\mu G$ and direction spatially correlated with the spiral arms (\cite{sofue, beck1996, wielebinski}).
For our Galaxy, the magnetic field direction also seems to follow the spiral arms but with a complex spatial distribution (\cite{wielebinski, han2006, beck2006}).Indeed, there are hints for local inversion of the field direction and radial dependency of the intensity (\cite{han2006, beck2001}). Pulsar Faraday rotation measurements (\cite{han2004,han2006, sofue, brown}) have been used to fit the Galactic large-scale magnetic field with various models including a axi-symmetric or bi-symmetric form, a field that reverses in the inter-arm regions, etc. Pulsar Faraday rotation measurements also indicate the presence of a turbulent component of the magnetic field (\cite{han2004}).\\

\subsubsection{Large-scale magnetic field.}

\indent In the following we consider a Modified Logarithmic Spiral (MLS) model of the large-scale magnetic field based on the WMAP team model presented in Page et al (2007). It assumes a logarithmic spiral to mimic the shape of the spiral arms (\cite{sofue}) to which we have added a vertical component. In galactocentric cylindrical coordinates $(r,z,\Phi)$ it reads

\begin{eqnarray}
\centering
\mathbf{B}(\mathbf {r})&=& B_{reg}(\mathbf {r})[ \cos(\phi+\beta) \ln \left( \frac{r}{r_0} \right) \sin(p) \cos(\chi ) \cdot \mathbf{u_r} \nonumber \\&&- \cos(\phi+\beta) \ln \left( \frac{r}{r_0} \right) \cos(p) \cos(\chi) \cdot \mathbf{u_{\phi}}  \nonumber \\&&+ sin(\chi) \cdot \mathbf{u_z}] ,
\end{eqnarray}
\noindent where $p$ is the pitch angle and  $\beta=1/\tan(p)$. $r_0$ is the radial scale and $\chi(r) = \chi_0(r)(z/z_0)$ is the vertical scale. 
Following \cite{taylor} we restrict our model to the range $3 < r < 20$ kpc. The lower limit is setted to avoid the center of the Galaxy
for which the physics is poorly constrained and the model diverges. The intensity of the regular field is fixed using pulsar Faraday rotation measurememts by \cite{han2006}
\be
B_{reg}(r) = B_0 \ e^{-\frac{r-R_{\odot}}{R_B}}
\ee

\noindent where the large-scale field intensity at the Sun position is $B_0=2.1 \pm 0.3 \mu
G$ and $R_B = 8.5 \pm 4.7 kpc$. The distance
between the Sun and the Galactic center, $R_{\odot}$ is set to 8 kpc (\cite{eisenhauer, reid}). \\

We also study the spiral model of \cite{stanev,sun}, hereafter ASS. In cylindrical coordinates it is given by
\begin{eqnarray}
\centering
B^D_r &=& D_1(r,\Phi,z)D_2(r,\Phi,z)sin(p)\\
B^D_{\Phi} &=& -D_1(r,\Phi,z)D_2(r,\Phi,z)cos(p)\\
B^D_z &=& 0
\end{eqnarray}
where $D_1$ accounts for  the spatial variations of the field and $D_2$
for asymmetries or reversals in the direction. The pitch angle is defined as for
the MLS model described above. $D_1 (r,z)$ is given by 

\begin{equation}
D_1(r,z) = \begin{cases}
B_0 \ exp(\frac{r-R_{\odot}}{R_0}- \frac{|z|}{z_0}) & r > R_c \\
B_c & r \leq R_c
\end{cases}.
\end{equation}
where $R_{\odot}$ is the distance of the Sun to the center of the Galaxy and it is set  to 8 kpc as before. $R_c$ is a critical radius
and it is set to 5 kpc following the ASS+RING model in \cite{sun}. In the same way $R_0$ is fixed to 10 kpc,
$B_0$ to 6 $\mu$G and $B_c$ to 2 $\mu$G. The field reversals are specified by
\begin{equation}
D_{2} (r) = \begin{cases}
+1 &  r  > 7.5  \\
-1 &  6 < r \le 7.5 \\
+1 &  5 < r \le 6 \\  
-1 &  r < 5
\end{cases}
\end{equation}
where $r$ is defined in units of kpcs. \\
It is important to notice that the synchrotron and thermal dust polarized emission depends only
on the orientation and not on the sign of the magnetic field and therefore, they are not sensible to field reverses in the model.

\subsubsection{Turbulent component}

\indent  In addition to the large-scale Galactic magnetic field, Faraday rotation measurements on pulsars in our vicinity have revealed a turbulent component for scales smaller than few hundred pc (\cite{lyne}). Moreover this turbulent field seems to be present at large scales (\cite{han2004}). The amplitude of this turbulent component is estimated to be of the same order of magnitude as the amplitude of the regular one (\cite{han2006}). 
The magnetic energy $E_B(k)$ associated with the turbulent component is well described by a Kolmogorov spectrum (\cite{han2004, han2006})
\be       
\centering
E_B(k) = C \left(\frac{k}{k_0}\right)^{\alpha}
\label{eq_pw_ko}
\ee
\noindent where $\alpha = -0.37$ and $C = (6.8 \pm 0.3) \cdot 10^{-13}\,\mathrm{erg\,cm^{-3}\,kpc}$.

\subsubsection{Final model}

Finally the total magnetic Galactic field $B_{tot}(\mathbf {r})$ can be written as
\be
\centering
B_{tot}(\mathbf {r}) = B_{reg}(\mathbf {r}) + B_{turb}(\mathbf {r})
\ee 
where $ B_{reg}(\mathbf {r})$ is the regular component, either MLS or ASS, and $B_{turb}(\mathbf {r})$ is the turbulent one.
We define $A_{turb}$ as the relative intensity of the turbulent component with respect to 
the regular one and it is a free parameter of the model. The turbulent component is computed from a 3D random realization
of the Kolmogorov spectrum presented above over a box of 512$^3$ points of 56 pc resolution.\\

In this paper we do not consider the halo component presented by \cite{sun,jansson}) as relativistic electrons and
dust grains are not expected to be concentrated on the halo.

\subsection{Emissivity model in polarization}
\label{buil_map}

As discussed in the previous section, the polarized emission in the 23~GHz WMAP data shows
complex structures both on the Galactic plane and in local high Galactic latitude structures such as
the North Celestial Spur (\cite{page2007}). An accurate representation of this complexity can
not be achieved using a simplified model like the one presented here. A similar
degree of complexity is observed in the 353~GHz polarization maps although the morphology
of the structures is rather different. To account for this complexity the Q and U estimate for
synchrotron and thermal dust models are corrected using intensity templates of these components 
extrapolated to the observation frequencies (23 and 353~GHz) using constant spectral
indeces.

\indent For the synchrotron emission we have

\begin{eqnarray}
\centering
Q_s &=& I_{\mathrm{Has}}  \left(\frac{\nu}{0.408}\right)^{\beta_s}\frac{Q^{sync}_{\nu}}{I^{sync}_{\nu}} ,\\
U_s &=& I_{\mathrm{Has}} \left(\frac{\nu}{0.408}\right)^{\beta_s}\frac{U^{sync}_{\nu}}{I^{sync}_{\nu}},
\end{eqnarray}

\noindent where $I_{\mathrm{Has}}$ is the reference map in intensity
constructed from the 408 MHz all-sky survey (section \ref{data_has}) after subtraction of the free-free emission
and $\nu$ is the frequency of observation.  Notice that we do not use the synchrotron MEM intensity map at 23~GHz (\cite{hinshaw}) as a synchrotron template to avoid any possible spinning dust contamination.
The spectral index $\beta_s$ used to extrapolate maps at various frequencies is assumed
to be spatially constant on the sky and is a free parameter of the model.

\indent For the thermal dust emission we write
\begin{eqnarray}
\centering
Q_d &=& I_{sfd}  \frac{Q^{dust}_{\nu}}{I^{dust}_{\nu}} ,\\
U_d &=& I_{sfd} \frac{U^{dust}_{\nu}}{I^{dust}_{\nu}} ,
\end{eqnarray}

where  $I_{sfd}$ is the reference map in intensity at 353 GHz
generated using model 8 of Schlegel, Finkbeiner and Davis
(\cite{finkbeiner}) which was obtained from the IRAS ~\cite{schlegel} and COBE/DIRBE data
~(\cite{boulanger}).

\indent We compute the above I,Q and U maps for synchrotron and thermal dust with the help of a modified version of 
the Hammurabi code (\cite{waelkens}). Each map is generated by integrating in 100 steps along each line-of-sight defined 
by the HEALPix $N_\mathrm{side}=128$ pixel centres.  The integration continues out to 25~kpc from the observer situated 
8.5~kpc from the Galactic centre.

\begin{table*}
\begin{center}
\caption{Latitude and longitude bands for the Galactic profiles used in the analysis.\label{gal_band}}
\vspace{0.3cm}
\begin{tabular}{|c|c|c|c|c|c|c|c|} \hline
Latitude interval (deg) & $[0,30]$& $[30,90]$& $[90,120]$ & $[120,180]$ & $[180,270]$ &
$[270,330]$& $[330,360]$ \\\hline
Longitude interval (deg) & $[-90,-50]$& $[-50,-20]$& $[-20,-5]$ & $[-5,5]$ & $[5,50]$ &
$[50,70]$& $[70,90]$ \\\hline
\end{tabular}
\end{center}
\end{table*}

\begin{table*}
\begin{center}
\caption{Parameters of the 3D Galactic model.\label{param_tab}}
\vspace{0.3cm}

\begin{tabular}{|c|c|c|} \hline
 Parameter  &   Range   &   Binning  \\\hline
$p$ (deg) &    $[-80.0,80.0]$  &   $10.0$   \\\hline
$A_{turb}$  &   $[0,2.5]*B_{reg}$  &   $0.25$ \\\hline 
$n_{\mathrm{CRE},r}$ (kpc)& $[0.0,10.0]$  &   $1$ \\\hline
$\beta_s$ & $[-4.3,-2.4]$  & $0.1$ \\\hline

\end{tabular}
\end{center}
\end{table*}


\section{Galactic-profiles comparison}
\label{gal_comp}

\subsection{Galactic-profiles description}

\begin{figure*}
\centering
\includegraphics[height=15cm,width=13cm]{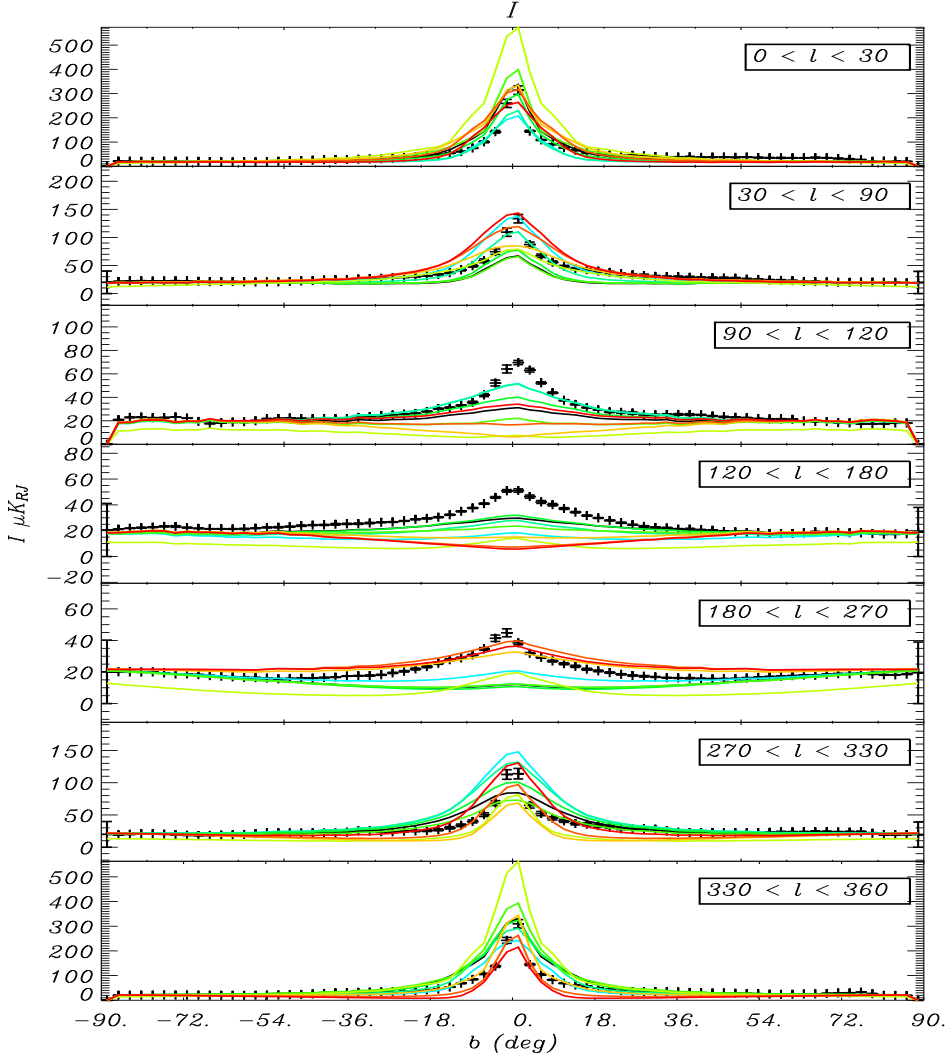}\caption{Galactic profiles in temperature at 408 MHz built using the Haslam data \emph{(black)} and the model of synchrotron emission with MLS for various values of the pitch angle $p$ \emph{(form green to red)}. \label{galprofil_has_BSS}}
\end{figure*}

\begin{figure*}
\centering
\includegraphics[height=15cm,width=7cm]{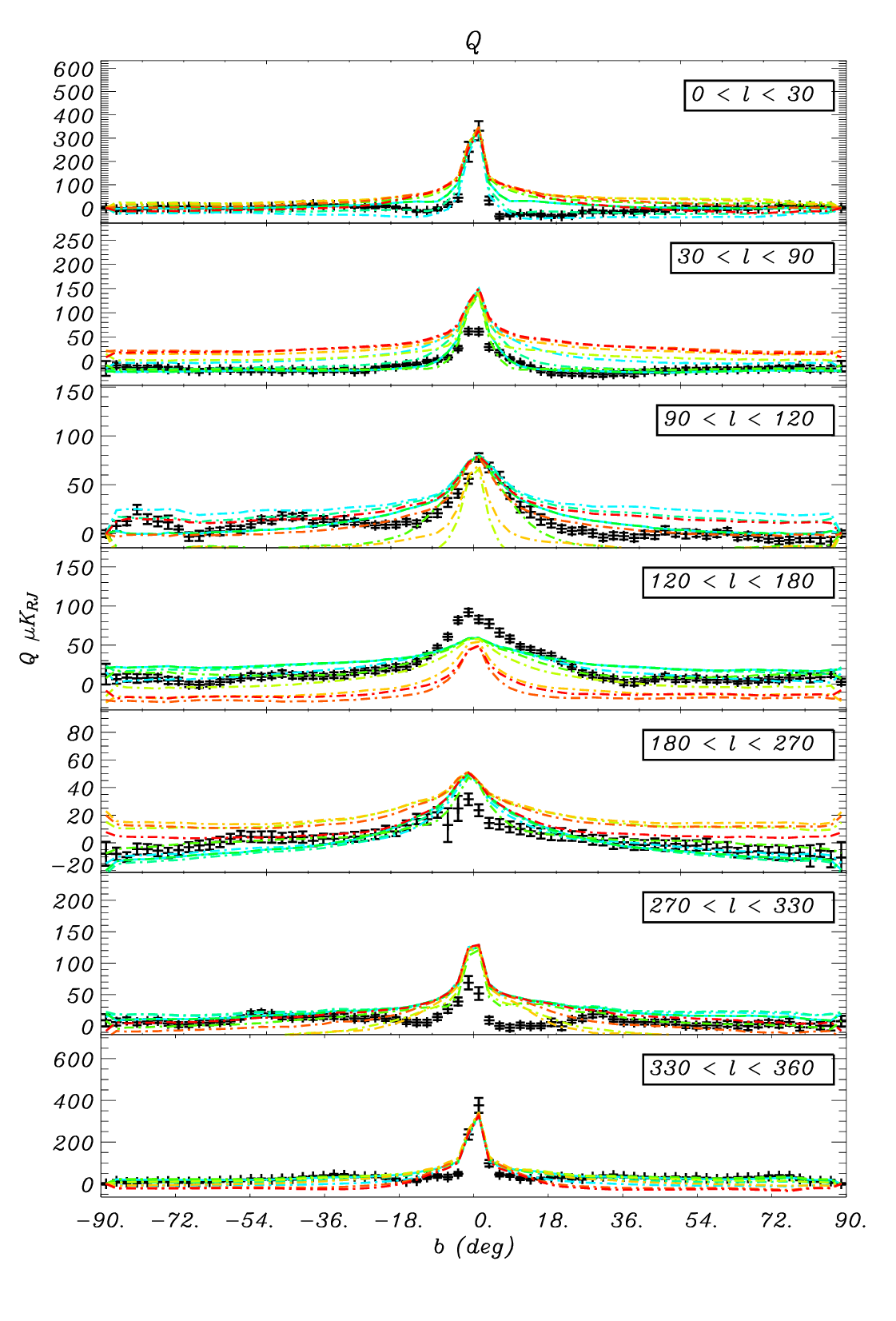}\includegraphics[height=15cm,width=7cm]{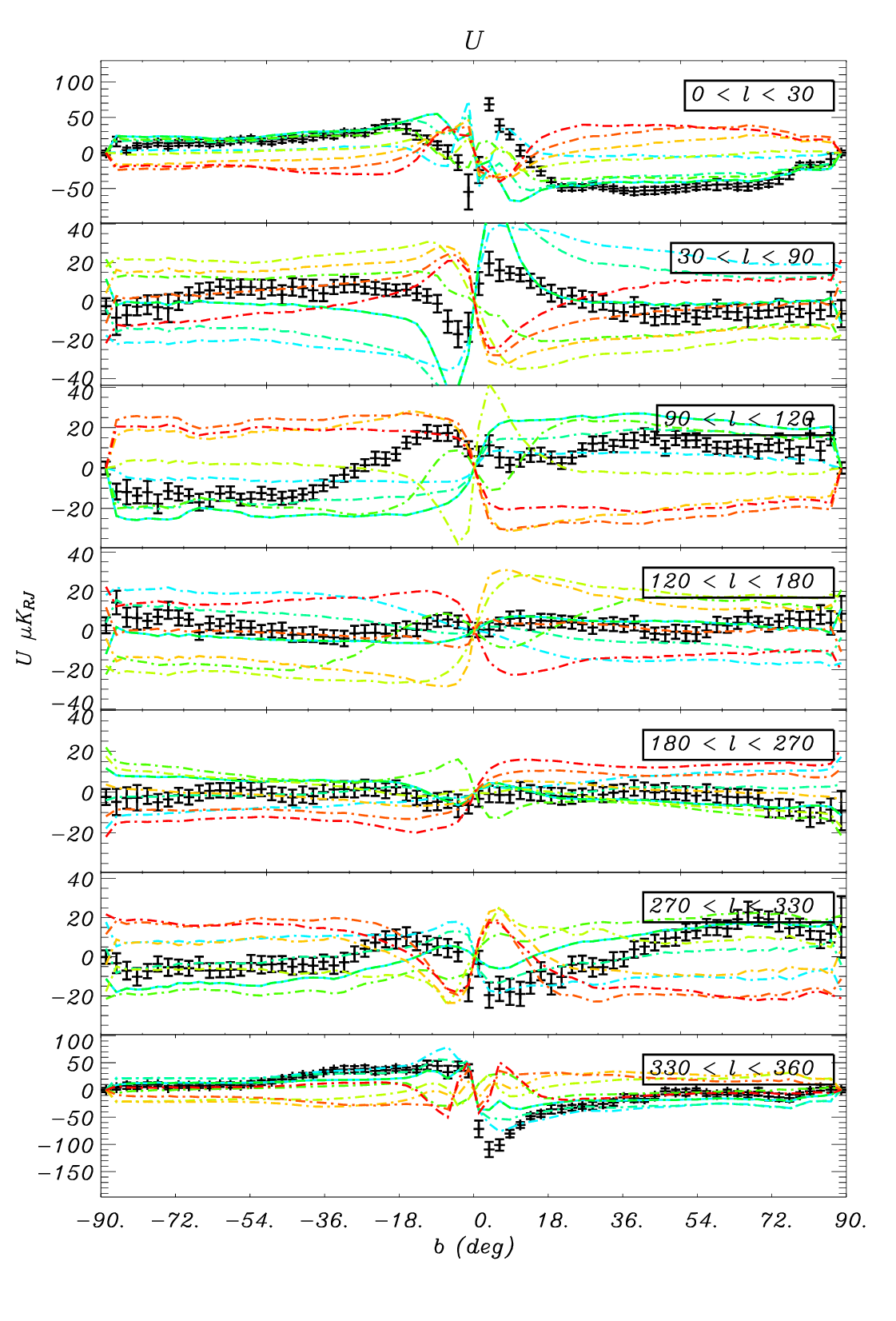}\caption{Galactic profiles in polarization Q and U at 23 GHz built with the five-year WMAP data \emph{(black)} and the model of synchrotron emission with MLS magnetic field for various values of the pitch angle $p$ \emph{(form green to red)}. \label{galprofil_wmap_BSS}}
\end{figure*}


\begin{figure*}
\centering
\includegraphics[height=15cm,width=7cm]{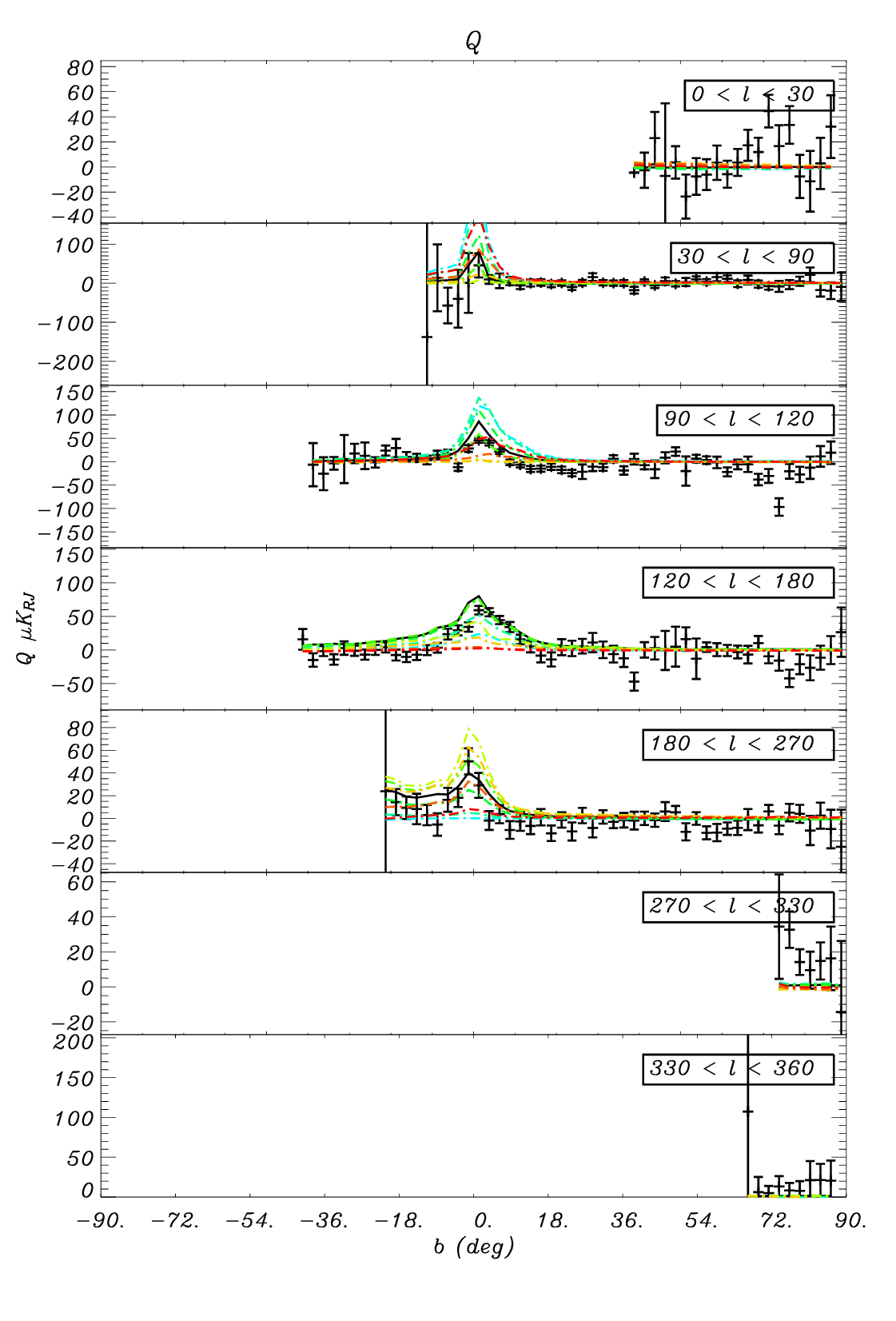}\includegraphics[height=15cm,width=7cm]{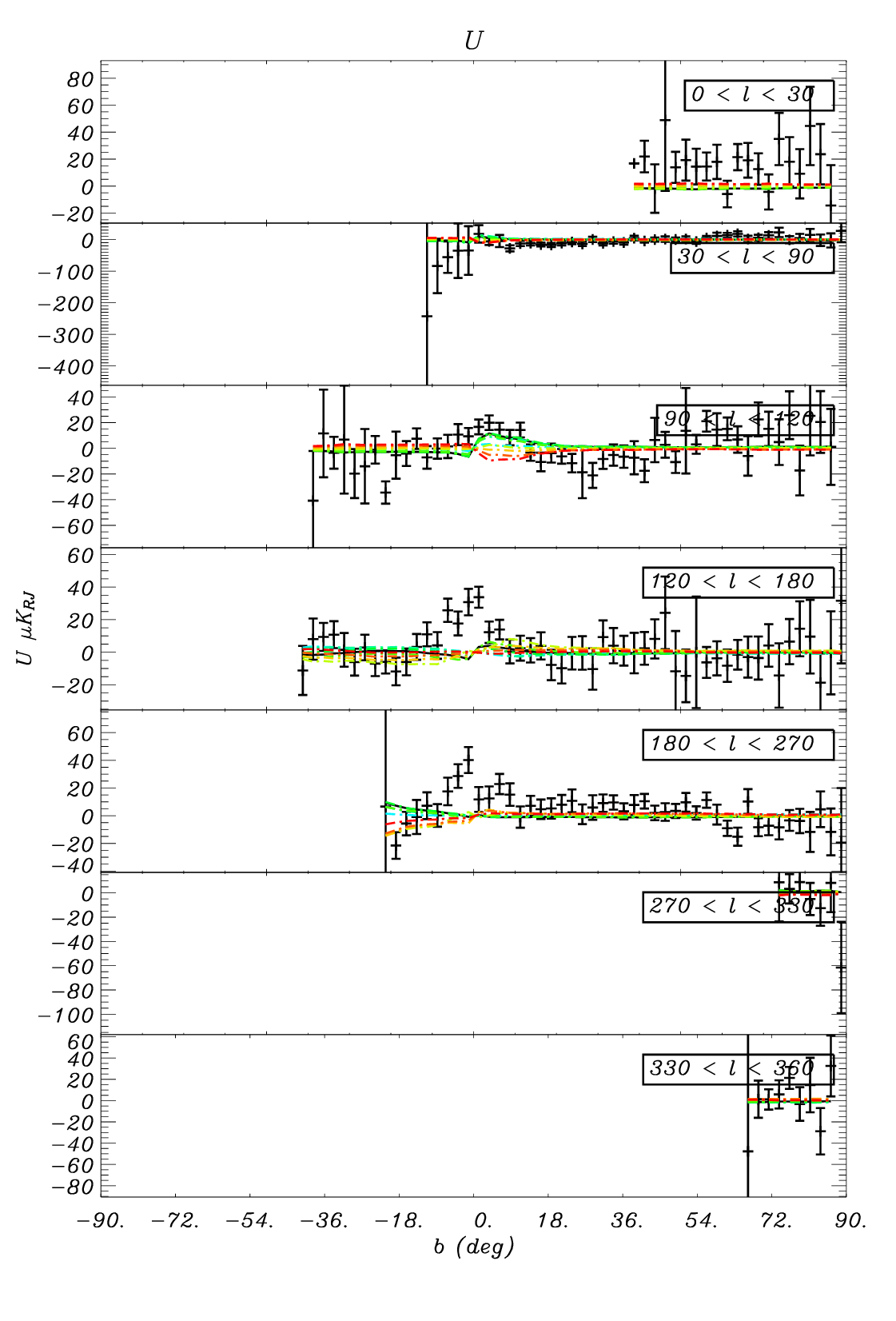}
\caption{Galactic profiles in temperature and polarization Q and U at 353 GHz with the ARCHEOPS data \emph{(black)} and for various values of the pitch angle $p$ for the model in polarization of thermal dust emission with MLS magnetic field \emph{(from green to red)}.\label{galprofil_arch_BSS}}
\end{figure*}

\indent In order to compare the models of Galactic polarized emission to the available data
we compute Galactic longitude and latitude profiles for the models and for the data in temperature and polarization
using the set of latitude and longitude bands defined in Table~\ref{gal_band}. For the Galactic longitude and latitude profiles both, we use
bins of longitude of $2.5^{\circ}$.
In the following discussions we will only consider Galactic latitude profiles because equivalent results are obtained
with the longitudinal profiles.  \\

We compute error bars including intrinsic instrumental uncertainties and the extra
variance induced by the presence of a turbulent component. The latter is estimated from the RMS within each of the latitude bins
following \cite{jansson}. For the 408~MHz all-sky survey we account for intrinsic uncertainties due to the 10 \% calibration errors
described in Section~\ref{data}. For the WMAP 23~GHz data we have computed 600 realizations of Gaussian noise maps from the number of hits per pixel
and the sensitivity per hit given on the WMAP LAMBDA web site. We have computed Galactic latitude
profiles in polarization for these simulated maps and estimated intrinsic errors from the standard deviation within each latitude bin. For the
Archeops data we use the noise simulations discussed in~\cite{macias} and proceed as for the WMAP data. \\

Galactic latitude profiles are computed for a grid of models obtained by varying the pitch angle, $p$,
the turbulent component amplitude, $A_{turb}$, the radial scale for the synchrotron emission,  $n_{\mathrm{CRE},r}$ and the synchrotron
spectral index, $\beta_s$. The range and binning step cosidered for each of these parameters is given in Table~\ref{param_tab}. 
All the other parameters of the model of the Galactic magnetic field and matter density are fixed to values proposed
in Section~\ref{3dgal_model}. Notice that to be able to compare the dust models to the Archeops 353~GHz data, the simulated maps
are multiplied by a mask accounting for the Archeops incomplete sky coverage. \\

Figure~\ref{galprofil_has_BSS} shows in black Galactic latitude profiles in temperature for the 408 MHz all-sky survey with error
bars computed as discussed above. In colors, we show for comparison the expected synchrotron diffuse Galactic emission from the MLS Galactic magnetic
field model for various values of the pitch angle $p$. On Figure~\ref{galprofil_wmap_BSS} we present the polarization Galactic
latitude profiles for the WMAP 23~GHz data (black) and the expected polarized diffuse synchrotron emission for the previous MLS models (color).
Finally, Figure~\ref{galprofil_arch_BSS} shows the polarization Galactic latitude profiles for the 353~GHz Archeops data (black) compared to
the same MLS models (color). From these figures we can see that the current available data do help discriminating between the different
models and therefore a likelihood analysis is justified.

\subsection{Likelihood analysis}
\label{result}

\begin{table*}
\begin{center}
\caption{Best-fit parameters for the MLS and ASS models of the Galactic magnetic field.\label{param}}
\vspace{0.3cm}

\begin{tabular}{|c|c|c|c|c|c|c|} \hline
Data  & Magnetic field model  & $ p (deg)$& $A_{turb} $  & $n_{\mathrm{CRE},r}$ & $\beta_s$ & $\chi^2_{min}$  \\\hline
408 MHz  & MLS   & $ -20.0^{+60.0}_{-50.0}$ & $< 1.00$  (95.4 \% CL)& $
4^{+16}_{-3} $ & $\emptyset$ & $3.58$     \\\cline{2-7}
        & ASS   & $-10.0^{+80.0}_{-70.0} $    & $ < 1.0$ (95.4 \% CL) &
        $5^{+15}_{-3}$ & $\emptyset$ & $4.65$ \\\hline
WMAP 23 GHz   & MLS & $ -30.0^{+40.0}_{-30.0}$ & $< 1.25$ (95.4 \% CL) & $ < 20$ (95.4
\% CL) &  $-3.4^{+0.1}_{-0.8}$ & $5.72$     \\\cline{2-7}
        &  ASS   & $-40.0^{+60.0}_{-30.0}$  & $ < 1.5$ (95.4 \% CL)   &  $ 3^{+17}_{-2}$(95.4\% CL) & $-3.4^{+0.1}_{-0.8} $ & $7.62$ \\\hline
Archeops 353 GHz  & MLS   & $ -20^{+80}_{-50}$   & $ < 2.25 (95.4 \% CL)$ &
$\emptyset$ & $\emptyset$ & $ 1.98 $   \\\cline{2-7}
         & ASS   & $60.0^{+20}_{-40}$    & $ 0.25^{+2.0}_{-0.25} $ & $\emptyset$ & $\emptyset$ & $1.72$   \\\hline
All  & MLS   & $ -20^{+80}_{-50}$   & $ < 2.25 (95.4 \% CL)$ &
$\emptyset$ & $\emptyset$ & $ 1.98 $   \\\cline{2-7}
         & ASS   & $60.0^{+20}_{-40}$    & $ 0.25^{+2.0}_{-0.25} $ & $\emptyset$ & $\emptyset$ & $1.72$   \\\hline

\end{tabular}
\end{center}
\end{table*}

The data and model Galactic latitude profiles are compared using a likelihood analysis where
the total likelihood function is obtained from
\be
\mathcal{L}_{tot} = {\Pi}_{d=1}^{3}  \mathcal{L}_{d}
\ee
where for each of the 3 data sets described above the log-likelihood function is given by

\be
- \log \mathcal{L}_{d} = \sum_{i} \sum_{j=0}^{N_{lon}-1} \sum_{k=0}^{N_{lat}-1} \frac{(D_{i,j,k}^{d}-M_{i,j,k}^{d})^2}{{\sigma_{i,j,k}^{d}}^2}
\ee

\noindent where $i$ represents the polarization state meaning intensity only for the 408 MHz all-sky survey, and, Q and U polarization for
the 23~GHz WMAP and 353~GHz Archeops data. $j$ and $k$ represents the longitude bands and latitude bins respectively. $D_{i,j,k}^{d}$ and $M_{i,j,k}^{d}$
corresponds to the data set $d$ and model for the $i$ polarization state, $j$ longitude band and $k$ latitude bin, respectively. $\sigma_{i,j,k}^{d}$
is the error bar associated to $M_{i,j,k}^{d}$. \\

\indent Table~\ref{param} presents the best--fit parameters for the three-data sets described above
and also for the combination of three of them (labeled All in the table). Results are presented
both for the MLS and ASS models of the Galactic magnetic field. The best-fit value
for the pitch angle, $p$, are in agreement within 1-$\sigma$ error bars for the three data sets
but for the Archeops data in the case of an ASS model. Notice that
our results are compatible with the pitch angle values presented in \cite{sun,page2007,dusta}.
The relative amplitude of the turbulent component, $A_{turb}$, is poorly constrained and
the data does not seem to favour a strong turbulent component either in the case of MLS or ASS models.
However, our results are compatible with the ones presented in \cite{sun,dusta,han2004, han2006} at the 2-$\sigma$ level.
The electronic density radial scale, $n_{\mathrm{CRE},r}$, is poorly constrained by the data both for MLS and ASS models
although our results are compatible with those of \cite{sun}. We also tested the possibility of a local contribution
to the electronic density as proposed by~\cite{sun}. We found that adding this local component does not improve either
the fit or the constraint on the radial scale. The best-fit value for the spectral index of the synchrotron
emission seems to be significantly lower than the one in \cite{sun,page2007}. This may due to
differences in the intensity template. Notice that we rescale the polarization intensity using the 408~MHz all-sky
survey to obtain a more realistic model.

\subsection{Temperature and polarization angular power spectra}
\begin{figure*}
\centering
\includegraphics[height=13cm,width=15cm]{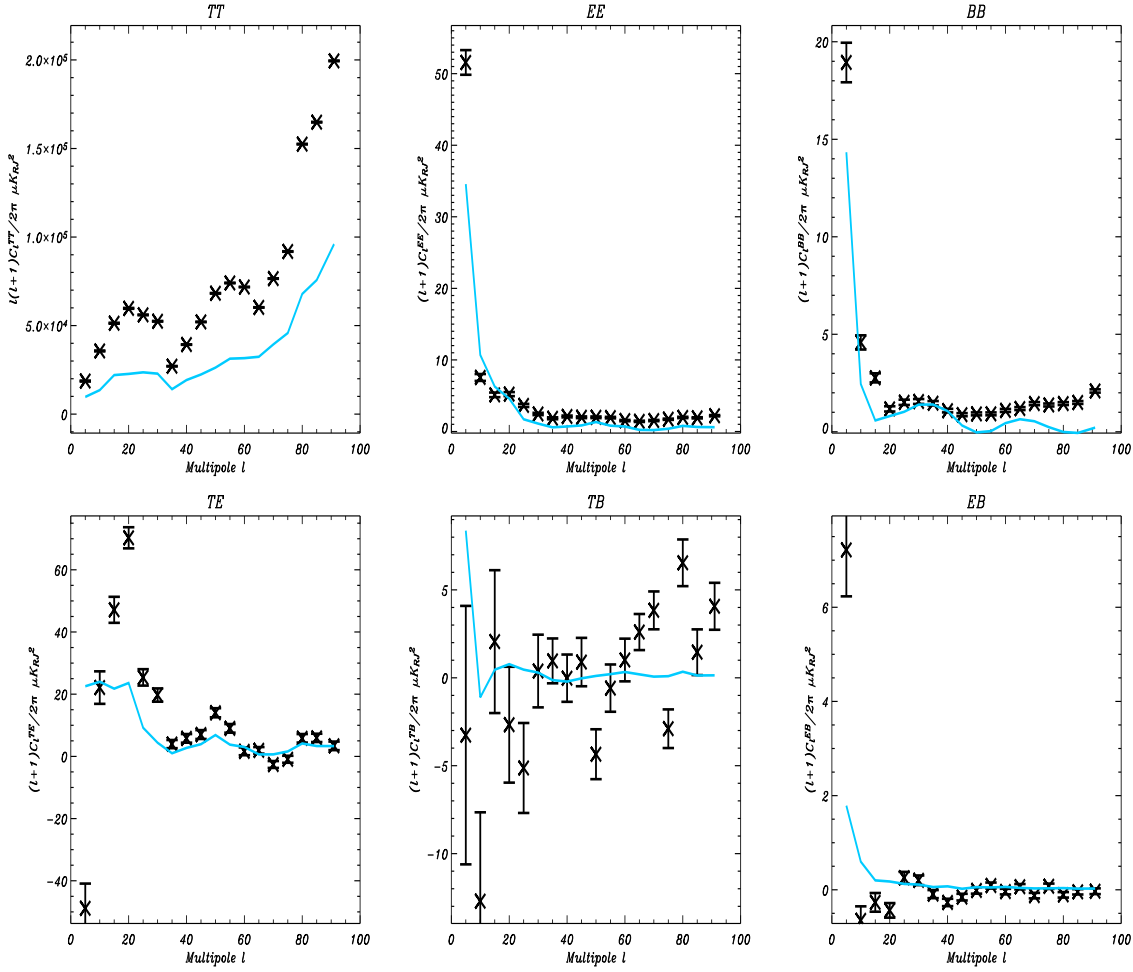}\caption{From left to right and from top to bottom : power spectra $C^{TT}_l$,$C^{EE}_l$,$C^{BB}_l$,$C^{TE}_l$,$C^{TB}_l$,$C^{EB}_l$ at 23 GHz built with the WMAP 5-year data \emph{(black)} and the model of synchrotron emission with MLS magnetic field for the best fit model parameters, excluding the Galactic region defined by $|b|<5^{\circ}$. \label{spec_wmap_gp}}
\end{figure*}

\begin{figure*}
\centering
\includegraphics[angle=90,height=13cm,width=15cm]{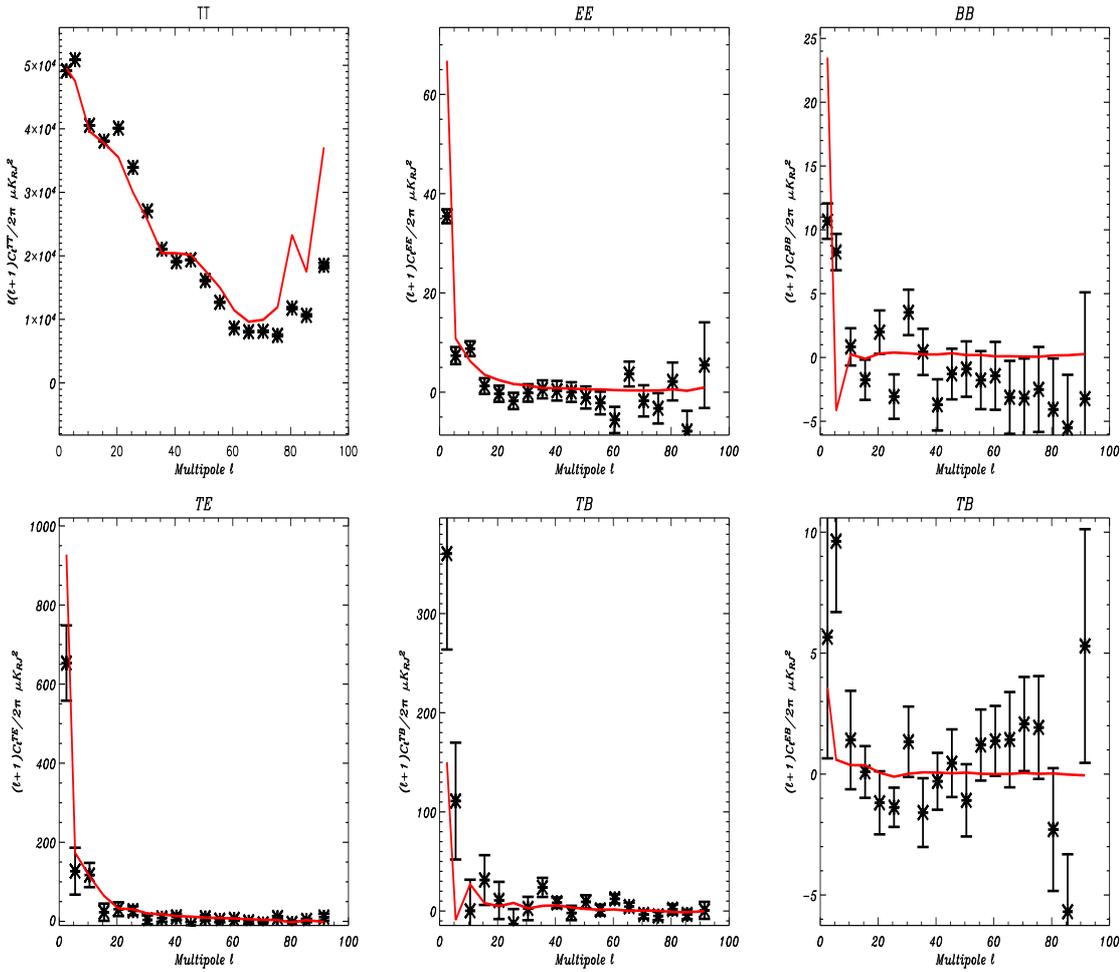}\caption{From left to right and from top to bottom: power spectra $C^{TT}_l$,$C^{EE}_l$,$C^{BB}_l$,$C^{TE}_l$,$C^{TB}_l$,$C^{EB}_l$ at 353 GHz computed from ARCHEOPS data \emph{(black)} and the model of thermal dust emission with MLS magnetic field for the best fit model parameters \emph{(red)} for the full-sky.\label{spec_dust_gp}}
\end{figure*}

\indent Using the best-fit parameters of the MLS model,
$p=-30.0^{\circ}$, $A_{turb} = 0.0$, $n_{\mathrm{CRE},r}$ = 1 and $\beta_s = -3.4$,
we have constructed simulated maps of the sky at 408 MHz and 23 and 353~GHz.
These maps are shown on right-hand side of Figures~\ref{data_has} and \ref{map_wmap_arch}.
Although for the Galactic profiles the fit can be considered relatively good, 
the fake temperature map at 408~MHz looks very different from the 408~MHz
all-sky survey map (left side of the plot), in particular at the North Polar Spur (\cite{wolleben07}),
as no local structures were included in the model. This supports a posteriori our correction
of the polarization synchrotron model using an intensity template as presented
in Section~\ref{3dgal_model}.
In Q and U polarization, the 23~GHz fake maps seem to reproduce
qualitatively the structure observed in the WMAP data (left side of the plot).
However in temperature the model and the data are very different as we have
not account for a variable synchrotron spectral nor fo any extra component as 
discussed in~\cite{page2007,kogut,dusta,macias2}.
Finally, the model of thermal dust emission is able to reproduce qualitatively the Archeops data
at 353 GHz. \\
\\
\indent Figures~\ref{spec_wmap_gp} and \ref{spec_dust_gp} show the temperature and polarization angular power spectra for the
23 GHz WMAP and 353~GHZ Archeops data compared to the best-fit MLS model for synchrotron and dust, respectively. 
As discussed before, the temperature auto power spectrum of the 23~GHz data is very different from the model as no extra
components in temperature were considered. However in polarization we have qualitatively a good agrement.
However, we clearly observe that the model does not account for all the observed emission. At 353~GHz the
agreement between the data and the model qualitatively and quantively is good. For polarization most of the
data samples at less than 3-$\sigma$ from the model. In temperature the model is not as accurate as in polarization
but we notice that the fitting was restricted to polarization data only. 

\section{Galactic foreground contamination to the CMB measurements by the Planck satellite}
\label{gal_bias}
\begin{figure*}
\centering
\includegraphics[height=13cm,width=15cm]{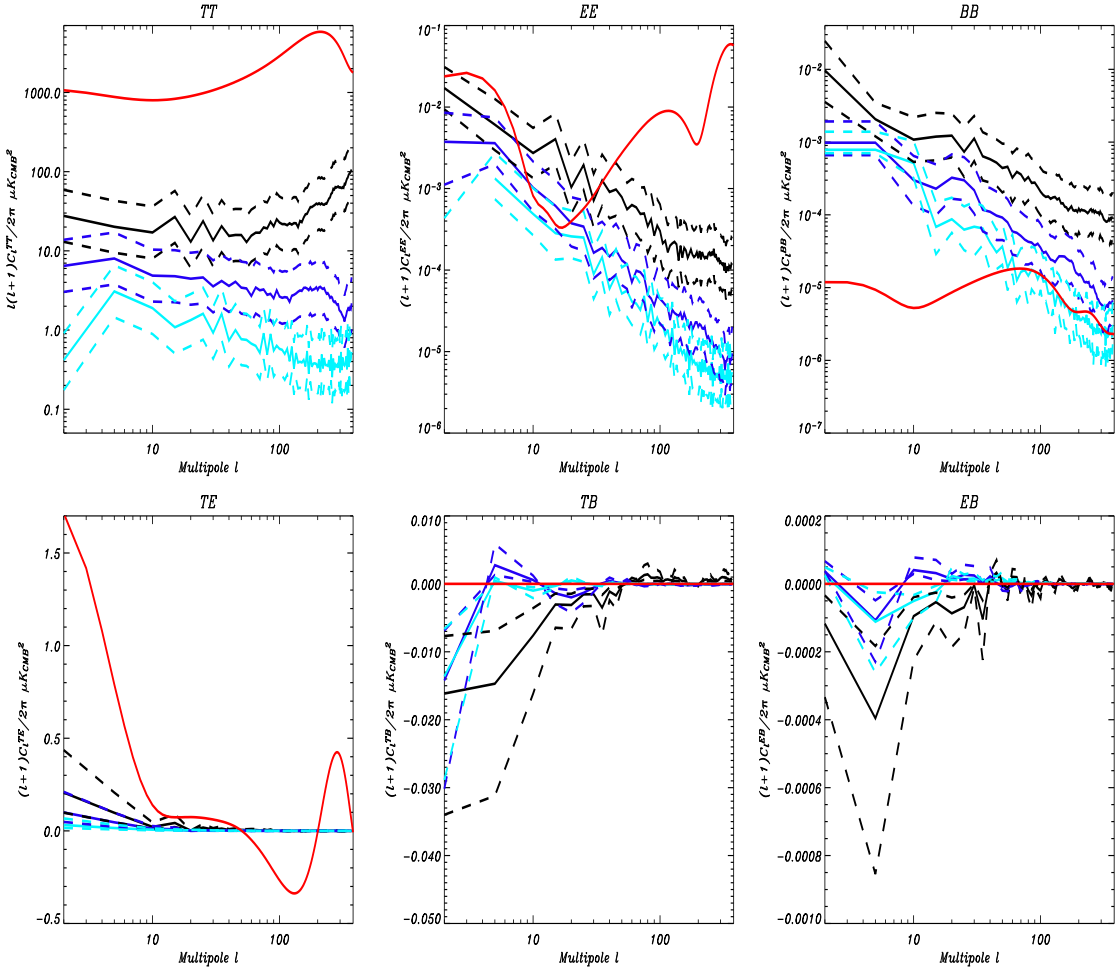}\caption{From left to right and from top to bottom: power spectra $C^{TT}_l$,$C^{EE}_l$,$C^{BB}_l$,$C^{TE}_l$,$C^{TB}_l$,$C^{EB}_l$ at 100 GHz the model of Galactic polarized emission applying a Galactic cut of $|b|<15^{\circ}$ (black)  $30^\circ$ (blue) and 
$40^\circ$ (cyan) and compared them 
to the expected CMB ones (red) for the WMAP best-fit $\Lambda$CDM model (\cite{komatsu2008}) to which we added a tensor component assuming an tensor to scalar ratio of 0.1..\label{spec_cmb}}
\end{figure*}

\indent  We can use the best-fit model of polarized synchrotron and dust emission
to estimate the  polarized foreground contamination to the CMB at the Planck satellite
observation frequencies. We are interested in comparing 
the expected foreground contribution to the expected CMB emission.
Notice that the aim of this section is not to obtain an accurate template of the polarized Galactic foreground emission to be subtracted on
the Planck data for CMB purposes.  \\

For this purpose, we have produced simulated maps of the Galactic polarized foreground emission using the best--fit model
parameters for each of the Planck CMB frequencies, 70, 100, 143 and 217~GHz. The thermal dust polarized emission have
been extrapolated using a constant spectral index of 2.0 in antenna temperature. We have computed the temperature
and polarization power spectra of these maps and compared them 
to the expected CMB ones for the WMAP best-fit $\Lambda$CDM model (available on the
LAMBDA website) to which we added a tensor component assuming a tensor to scalar ratio of 0.1.
Note that neither noise nor systematics nor resolution effects are considered.

Figure \ref{spec_cmb} shows these power spectra at 100 GHz. The expected CMB signal is represented in red. The polarized diffuse foreground
emission for Galactic latitude cuts of $|b| < 15^\circ$, $30^\circ$ and $40^\circ$ are shown in solid black, blue and cyan lines, respectively.
1-$\sigma$ errors in the model are represented are dashed lines. In temperature, the CMB $C^{TT}_l$ dominates at all the angular scales
considered as could be expected from the WMAP and Archeops data. For polarization, the CMB  $C^{EE}_l$ dominates at high $\ell$ values
but we observe significant foreground contamination at the lowest $\ell$ values ($\ell < 20$). In the same way, the CMB $C^{TE}_l$ 
dominates at 100~GHz but for very low $\ell$ values. However, the CMB $C^{BB}_l$ is significantly smaller than the foreground contribution
at all the angular scales considered even for such a large value of the tensor to scalar ratio. The CMB $C^{TB}_l$ and $C^{EB}_l$ are
expected for most cosmological models to be null and therefore, the foreground contribution dominates the signal.

As the Galactic polarized foreground emission seems to dominate the observed emission at the Planck CMB frequencies,
special care should be taken when estimating the CMB emission using standard template subtraction techniques and component separation
algorithms. The assesment of the final errors is crucial and we think that models of the polarized foreground emission such as those presented
in this paper can be of much help for this task.

\section{Summary and conclusions}
\label{conc}

\indent We have presented in this paper a detailed study of the
diffuse synchrotron and thermal dust polarized Galactic foreground
emission components.  We have constructed coherent models of these two
foregrounds based on a 3D representation of the Galactic magnetic
field and of the distribution of relativistic electrons and dust
grains in the Galaxy. For the Galactic magnetic field we have assumed
a large-scale regular component plus a turbulent one. The relativistic
electron distribution and dust grain distribution have been modeled as
exponentials peaking at the Galactic center.  From these analyses we
have been able to study the main parameters of the models, the
magnetic field pitch angle, $p$, the radial width of the relativistic
electron distribution, $h_{er}$, the relative amplitude of the
turbulent component, $A_{turb}$ and spectral index of the synchrotron
emission $\beta_{s}$. We have been able to set constraints only on the
pitch angle and the synchrotron spectral index. An upper limit on the
relative amplitude of the turbulent component is obtained although the
data seems to prefer no turbulence at large angular scales. With the
current data we are not able to constrain the radial width of the
relativistic electron distribution.  Notice that our constraints are
compatible with those in the literature.

\indent Using the best--fit parameters we have constructed
maps in temperature and polarization for the synchrotron and dust thermal
emission at 23 and 353~GHz and compare them to the WMAP and ARCHEOPS data
at the same frequencies. We find good agreement between
the data and the model. However, when comparing the temperature
and polarization power spectra for the data and model maps, we observe
that synchrotron emission model is not realistic enough. For dust the
model seems to reproduce better the data but it is important to realize that
the errors on the Archeops data are much larger.

\indent From this, we can conclude that the models presented in this
paper can not be used for direct subtraction of polarized foregrounds for CMB
purposes. However they can be of great help for estimating the impact of
the polarized Galactic foreground emission on the reconstruction of the CMB
polarized power spectra. Indeed, we have extrapolated the expected polarized
Galactic foreground emission to the Planck CMB frequencies, 70, 100, 143 and
217~GHz and found they dominate the emission at low $\ell$ values where 
the signature in the polarized CMB power spectra of important physical 
processes like reionization is expected. Furthermore, the Galactic polarized
foreground emission seems to dominate the B modes for which we expect an unique
signature from primordial gravitational waves. Because of this, we
propose the use of models like the ones presented in this paper to asses
the errors in the reconstruction of the CMB emission when using
template subtraction techniques or component separation algorithms.


\begin{thebibliography}{}
\bibitem[Battistelli et al~{2006}]{battistelli2006} Battistelli, E.S. et al, 2006, ApJ, {\bf 645}, 141-144
\bibitem[Baumann et al~{2009}]{baumann} Baumann~D. et al, 2009, AIP
  Conf.Proc., {\bf 1141}.
\bibitem[Beck et al~{1996}]{beck1996} Beck~R. et al, 1996, ARA\&A, {\bf 34}, 155.
\bibitem[Beck~{2001}]{beck2001} Beck~R., Space Science Reviews, {\bf 99}, 243.
\bibitem[Beck~{2006}]{beck2006} Beck~R., 2006, \emph{Proceedings of Polarization 2005}, EAS Publication Series
\bibitem[Beno\^it et al~{2004}]{benoit2004a} Beno\^it et al, 2004, A \& A, {\bf 424}, 571.
\bibitem[Betoule et al~{2009}]{betoules2009} Betoule et al, 2009, A \& A, {\bf 503}, 691B.
\bibitem[Boulanger et al~{1996}]{boulanger} Boulanger~F. et al, 1996, A\&A, {\bf 312}, 181.
\bibitem[PLANCK bluebook 2004]{bluebook} The Planck Consortia, \emph{The Scientific Program}, 2004. 
\bibitem[Brouw \& Spoelstra~{1976}]{brouw} Brouw~N.W. \&  Spoelstra~T.A.T., 1976, A.\& A. Sup. S., {\bf 26}, 129.
\bibitem[Brown et al~{2007}]{brown} Brown~J.C. et al., 2007, ApJ, {\bf 663}, 258-266.
\bibitem[Burn et al~{1966}]{burn1966} Burn~B. J. et al 1966, MNRAS, {\bf 133}, 67B.
\bibitem[Carretti et al~{2009}]{carretti2009} Carretti~E. et al 2009, {\bf astro-ph/0907.4861v1}
\bibitem[Cordes \& Lazio {2002}]{cordes} Cordes~J.M. \& Lazio T.J.W., 2002, {\bf astro-ph/0207156}.
\bibitem[Davis \& Greenstein~{1951}]{davis} Davis~B. T. \& Greenstein~J. L., 1951, ApJ, {\bf 114}, 206.
\bibitem[Desert et al~{1998}]{desert1998} Desert~F.-X. et al, 1998, A\&A, {\bf 342}, 363.
\bibitem[Desert et al~{2008}]{desert2008} Desert~F.-X. et al, 2008, A\&A, {\bf 481}, 411D.
\bibitem[Dickinson et al~{2003}]{dickinson} Dickinson~C., Davies~R. D. \& Davis~R. J., 2003, MNRAS, {\bf 341}, 369.
\bibitem[Drimmel \& Spergel~{2001}]{drimmel} Drimmel~R. \& Spergel~D.N., 2001, ApJ, {\bf 556}, 181.
\bibitem[Duncan~{1999}]{duncan1999} Duncan~A. et al, 1999, A. \& A., {\bf 350}, 447.
\bibitem[Eisenhauer et al~{2003}]{eisenhauer} Eisenhauer~F. et al, 2003, ApJ, {\bf 597}, L121.
\bibitem[Efstathiou et al~{2009}]{efstathiou1} Efstathiou~G. et al, 2009, MNRAS, {\bf 397}, 1355.
\bibitem[Efstathiou \& Gratton~{2009}]{efstathiou2} Efstathiou~G \& Gratton~S., 2009, JCAP,{\bf 6}, 11.
\bibitem[Finkbeiner et al~{1999}]{finkbeiner} Finkbeiner~D. P., Davis~M. \& Schlegel~D. J., 1999, ApJ, {\bf 524}, 867 .
\bibitem[Gold et al~{2009}]{gold} Gold~B. et al, 2009, ApJS, {\bf 180}, 265.
\bibitem[Goodman \& Whittet~{1995}]{goodman1995} Goodman A. A. \& Whittet D. C. B., 1995, ApJ, {\bf 455}, 181.
\bibitem[G\'orski et al~{2005}]{gorski} G\'orski~K.M. et al, 2005, ApJ, {\bf 622},759.
\bibitem[Hall~{1949}]{hall}Hall~J.S.,1949, Science, {\bf 109}, 106.
\bibitem[Han et al~{2004}]{han2004} Han~J. L., Ferri\`ere~K. \& Manchester~R. N., 2004, A\&A, {\bf 610}, 820-826.
\bibitem[Han et al~{2006}]{han2006} Han~J. L. et al, 2006, A\&A, {\bf 642}, 868.
\bibitem[Haslam et al~{1982}]{haslam} Haslam~C.G.T.et al, 1982, A\&AS, {\bf 47}, 1.
\bibitem[Heiles~{2000}]{heiles} Heiles~C.,2000, ApJ, {\bf 119}, 923.
\bibitem[Hildebrand et al~{1999}]{hildebrand1999} Hildebrand~R. H. et al, 1999, ApJ, {\bf 516}, 834.
\bibitem[Hiltner~{1949}]{hiltner} Hiltner~W.A., 1949, Science, {\bf 109}, 65.
\bibitem[Hinshaw et al~{2007}]{hinshaw} Hinshaw~G. et al, 2007, ApJS, {\bf 170}, 288.
\bibitem[Jaffe et al~{2009}]{jaffe} Jaffe~T. et al, 2009, MNRAS,{\bf 401}, 1013.
\bibitem[Jansson et al~{2009}]{jansson} Jansson~R. et al, 2009, JCAP, {\bf 7}, 21.
\bibitem[Kogut et al~{2007}]{kogut} Kogut~A. et al, 2007, ApJ, {\bf 665}, 355.
\bibitem[Komatsu et al~{2008}]{komatsu2008} Komatsu~E. et al, 2008, ApJS, {\bf 180}, 306.
\bibitem[Lazarian~{1995}]{lazarian1995} Lazarian~A., 1995, MNRAS, {\bf 277}, 1235-1242.
\bibitem[Lazarian~{1997}]{lazarian1997} Lazarian~A. et al, 1997, ApJ, {\bf 490}, 273.
\bibitem[Lazarian~{2009}]{lazarian2009} Lazarian~A. et al, 2009, ASP Conference Series, {\bf 4}.
\bibitem[Leach et al~{2008}]{leach} Leach~S. M. et al, 2008, A\&A, {\bf 491}, 597.
\bibitem[Lyne \& Smith~{1989}]{lyne} Lyne~A.G. \& Smith~F.G., 1989, MNRAS, {\bf 237}, 533.
\bibitem[Mathis~{1986}]{mathis1986} Mathis J. S., 1986, ApJ, {\bf 308}, 281. 
\bibitem[Mac\'ias-P\'erez et al~{2007}]{macias} Mac\'ias-P\'erez~J. F., Lagache~G., Maffei~F. et al.,  2007, A\&A, {\bf 467} 1313.
\bibitem[Mac\'ias-P\'erez et al~{2010}]{macias2} Mac\'ias-P\'erez~J. F.,D\'esert~F.-X., Tristram~M., Fauvet~L. et al in preparation.
\bibitem[Miville-Desch\^enes et al~{2008}]{dusta} Miville-Desch\^enes~M. -A. et al, 2008, A\&A, {\bf 490}, 1093.
\bibitem[Nolta et al~{2009}]{nolta2009} Nolta~M.R., 2009, ApJS, {\bf 180}, 296.
\bibitem[Page et al~{2003}]{page2003} Page~L. et al, 2003, ApJS, {\bf 148}, 39.
\bibitem[Page et al~{2007}]{page2007} Page~L. et al, 2007, ApJSS, {\bf 170}, 335.
\bibitem[Peiris et al~{2003}]{peiris} Peiris~H. et al, 2003, ApJS, {\bf 148}, 213.
\bibitem[Ponthieu et al~{2005}]{ponthieu2005} Ponthieu~N., Mac\'ias-P\'erez~J. F, Tristram~M. et al, 2005, A\&A, {\bf 444}, 327.
\bibitem[Reid \& Brunthaler~{2005}]{reid} Reid~M. \& Brunthaler~A., 2005, in ASP Conf. Ser. 340, Future Directions in High Resolution Astronomy: The 10th Anniversary of the VLBA, ed. J. Romney \& M. Reid (San Fransisco: ASP), 253 .
\bibitem[Ribicki \& Lightman~{1979}]{ribicki} Ribicki~G.B. \& Lightman~A., 1979, Radiative Process in Astrophysics (New York, Wiley-Interscience). 
\bibitem[Schlegel et al~{1998}]{schlegel} Schlegel~D. J., Finkbeiner~D.P. \& Davis~M., 1998, ApJ, {\bf 500}, 525.
\bibitem[Sofue et al~{1986}]{sofue} Sofue~Y., Fujimoto~M. \& Wielebinski~R., 1986, ARA\&A, {\bf 24}, 459.
\bibitem[Stanev~{1997}]{stanev} Stanev~T., 1997, ApJ, {\bf 479}, 290.
\bibitem[Sun et al~{2008}]{sun} Sun~X.H., Reich~W., Waelkens~A. \& Ensslin~T.A., 2008, A\&A, {\bf 477}, 573.
\bibitem[Taylor \& Cordes~{1993}]{taylor} Taylor~J. H. J. \& Cordes~J. M., 1993, ApJ, {\bf 411}, 674.
\bibitem[Uyaniker et al~{1999}]{uyaniker} Uyaniker~B. et al, 1999, A\&AS, {\bf 132}, 401.
\bibitem[Vaillancourt~{2002}]{vaillancourt} Vaillancourt~J. E., 2002, ApJS, {\bf 142}, 335.
\bibitem[Waelkens et al~{2009}]{waelkens} Waelkens~A., Jaffe~T. et al, 2009 A\&A, {\bf 495}, 697.
\bibitem[Wielebinski~{2005}]{wielebinski} Wielebinski~R., 2005, \emph{Cosmic Magnetic Field}, ed. R. W. R. reck, Springer, Berlin.
\bibitem[Wolleben et al~{2006}]{wolleben} Wolleben~M. et al, 2006, A.\& A., {\bf 448}, 411. 
\bibitem[Wolleben et al~{2007}]{wolleben07} Wolleben~M. et al, 2007, ApJ, {\bf 664}, 349.

\end{thebibliography}
\end{document}